\documentclass[twocolumn,amssymb,fleqn]{revtex4} %fleqn: make all eqns left alighed  ,showpacs
\usepackage{epsfig,amssymb,amsmath,graphicx,subfigure,hyperref  }  % pdfpages and graphicx are added   hyperref(set hyperlink for content)
\usepackage{mathrsfs}
\usepackage{color}
\usepackage{multirow}
\usepackage{tabularx}

\begin{document}

\title{Two and three electrons on a sphere: A generalized Thomson problem} 
\author{Liu Yang and Zhenwei Yao} 
\email{zyao@sjtu.edu.cn}
\affiliation{School of Physics and Astronomy, and Institute of Natural Sciences, Shanghai
Jiao Tong University, Shanghai 200240, China} 
\begin{abstract}
Generalizing the classical Thomson problem to the quantum regime provides an ideal model to
explore the underlying physics regarding electron correlations. In this work, we
systematically investigate the combined effects of the geometry of the substrate and the
symmetry of the wave function on correlations of geometrically confined electrons. By
the numerical configuration interaction method in combination with analytical theory, we
construct symmetrized ground-state wave functions; analyze the energetics, correlations,
and collective vibration modes of the electrons; and illustrate the routine for 
the strongly correlated, highly localized electron states with the
expansion of the sphere. This work furthers our understanding about electron correlations
on confined geometries and shows the promising potential of exploiting
confinement geometry to control electron states. 
\end{abstract}

\maketitle

\section{Introduction}

Inquiry into the physics of geometrically confined electrons is a prominent research
theme in modern physics and chemistry~\cite{reimann2002electronic, sabin2009advances, koning2014crystals},
and it can be traced back to the classical problem of determining the ground state of
classical charged particles confined on the surface of a sphere, which is known as the
Thomson
problem~\cite{thomson1904xxiv,nelson2002defects,Bowick2002,levin2003charges,
bausch2003grain,de2005electromagnetic, wales2006structure, wales2009defect,
 mehta2016kinetic}.  The Thomson
problem and its various generalized versions arise in diverse physical
systems~\cite{dinsmore2002colloidosomes, Bowick2002, Bowick2006, cioslowski2009modified,
koning2014crystals, agboola2015uniform, 
yao2016dressed, yao2017topological, chen2018depletion}, ranging
from surface ordering of liquid-metal drops~\cite{davis1997}, colloidal
particles~\cite{bausch2003grain}, and protein subunits
over spherical viruses~\cite{Caspar1962,lidmar2003virus} to mechanical-instability-driven wrinkling crystallography
on spherical surfaces~\cite{brojan2015wrinkling}.
Recently, due to advances in semiconductor technology and spectroscopic probes,
geometrically confined few-electron systems have been experimentally accessible, and they
bring a host of scientific problems related to understanding electron
correlations~\cite{qdotRMP, artificial,quantum_disks,polygonal,ellipsoidal}.
Generalizing the classical Thomson problem to quantum regime provides an ideal model to
explore the underlying physics regarding electron correlations. In comparison with the
classical Thomson problem, its quantum version can exhibit richer physics beyond
minimization of Coulomb potential energy. For example, even a single electron will
interfere with itself when confined on the sphere. Furthermore, the Heisenberg uncertainty
principle requires that the electrons are always restless even in the ground state.

Past studies have shown the crucial role of system size on electron
states~\cite{correlation2,2el,excited3,nodal,2esph}. On large spheres, the confined
electrons become strongly correlated, which is closely related to Wigner crystallization
of uniform electron gas~\cite{crystal}.  Furthermore, studies of multiple-electron systems have revealed the fundamental role of symmetries of the wave function under rotation, inversion, and permutation and its non interaction feature 
on the nodal structure of electron states~\cite{nodal, loos2015nodal}. 
Electron states of two- or three-electron systems on the sphere have been extensively
studied using the approaches of approximate Schr\"{o}dinger equations~\cite{2el,2esph,
seidl2007adiabatic} and the configuration interaction (CI)
method~\cite{correlation2,excited3,2esph,helgaker2013molecular}. Notably, the system of
two electrons on a hypersphere has been quasiexactly solved, and the analytical results
are useful in the development of correlation functionals within density-functional
theory~\cite{loos2009two, loos2010excited, loos2011thinking}. In this work, we focus on
the combined effects of the geometry of the sphere and the symmetry of the wave functions
on energetics and correlations of confined electrons. The model of the quantum version
of the Thomson problem provides the opportunity to address fundamental questions with broader
implications, such as the following: How do the electrons become correlated with the expansion of the
sphere? What are the dynamic behaviors of the strongly correlated electrons?

To address these questions, we resort to the CI method in
combination with analytical theory to construct and analyze the ground state wave
functions of two- and three-electron systems~\cite{correlation2, 2esph}. Note that the CI
method allows us to analyze the variation of the components composing the ground-state
wave functions, and it provides insights into the enhancement of electron correlations. 
In this work, we construct symmetrized ground-state wave functions for both two- and
three-electron systems. Energetics analysis shows the degeneracy of wave functions with
distinct symmetries and the domination of the potential energy over the kinetic energy
with the expansion of the sphere. In this process, eigenstates with larger angular momentum
quantum numbers are excited under the increasingly important Coulomb interaction.
Consequently, strongly correlated, highly localized electron states are established in
the large-$R$ regime, as revealed in the probability analysis. In this regime, we propose
a semi classical small-oscillation theory to quantitatively analyze the vibration modes
and determine the symmetry-dependent quantum number of the ground-state harmonic
oscillations. The results presented in this paper further our understanding about
electron correlations in confined geometries and show the promising potential of
exploiting confinement geometry to manipulate electron states.

\section{Model and Method}

The ground-state wave function and energy of $N$ electrons on the sphere are determined by the
time-independent Schr\"{o}dinger equation:
\begin{eqnarray}
  \hat{H} \Psi(\{\vec{r}_i\}) =E\Psi(\{\vec{r}_i \}), \label{sch}
  \label{single_electron_1}
\end{eqnarray}
where $\hat{H}=\hat{K}+\hat{V}$ and $\vec{r}_i$ is the position of electron $i$. The
kinetic-energy term $\hat{K}=\sum_{i=1}^{N}\hat{L}_i^2/2R^2$, $R$ is the radius of the
sphere, and $\hat{L}_i$ is the angular momentum operator of the electron $i$. The potential
energy term $\hat{V} = \sum_{i<j} 1/|\vec{r}_i-\vec{r}_j|$.

We resort to the CI method to construct the ground
state wave functions with certain symmetries~\cite{helgaker2013molecular}. The CI wave function consists of a linear combination
of basis wave functions, whose expansion coefficients are variationally determined. This
method
can provide highly accurate wave functions, especially for systems with a small number of particles.
The CI method has extensive applications in quantum chemistry due to the simple structure
of the wave function~\cite{qchemistry}. In practice, a truncated Hilbert space spanned by dominant eigenstates
provides a good approximation for performing the diagonalization of the Hamiltonian.

We first construct the basis wave function $\Psi_{\bold{n}}(\{
  \vec{r}_i \})$, where $\bold{n}$ represents a complete set of quantum
numbers to characterize the state of the system. For the two-electron
system, $|\mathbf{n}\rangle =
|l_1,l_2,l,m\rangle$, which is the common eigenstate of $\hat{L}^2$, $\hat{L}_z$, and
$\hat{L}_1^2$. For the three-electron system, $|\mathbf{n}\rangle =
|l_{12},l_1,l_2,l_3,l,m\rangle$, which is the common eigenstate of the mutually commuting $\hat{L}_{12}^2$, $\hat{L}_{1}^2$, $\hat{L}_{2}^2$,
$\hat{L}_{3}^2$, $\hat{L}^2$, and $\hat{L}_z$. From the linear combination of $\Psi_{\bold{n}}(\{
  \vec{r}_i \})$, we construct wave functions $\Psi^{X_p}_{\bold{n}}(\{
  \vec{r}_i \})$ with certain symmetries. The superscript $X_p$ indicates that
the wave function is exchange-symmetric ($X=S$) or exchange-antisymmetric
($X=A$), and has even ($p=e$) or odd ($p=o$) parity. The relevant matrix
elements of the kinetic and potential energies are:
$K^{X_p}_{\bold{n}\bold{n'}}=\langle\Psi^{X_p}_{\bold{n}}|\hat{K}|\Psi^{X_p}_{\bold{n'}}\rangle$,
and
$V^{X_p}_{\bold{n}\bold{n'}}=\langle\Psi^{X_p}_{\bold{n}}|\hat{V}|\Psi^{X_p}_{\bold{n'}}\rangle$.

In this work, the units of length, energy and angular momentum are the Bohr
radius $a_{B}=4\pi\epsilon_0\hbar^2/m_e e^2$,
$e^2/4\pi\epsilon_0a_{B}$, and $\hbar$, respectively. $R/r_B \ll 1$ and $R/r_B
\gg 1$ in the small- and large-$R$ regimes, respectively.

\section{Results and discussion}

\begin{figure}
\centering
\subfigure[]
{\includegraphics[width=0.35\textwidth]{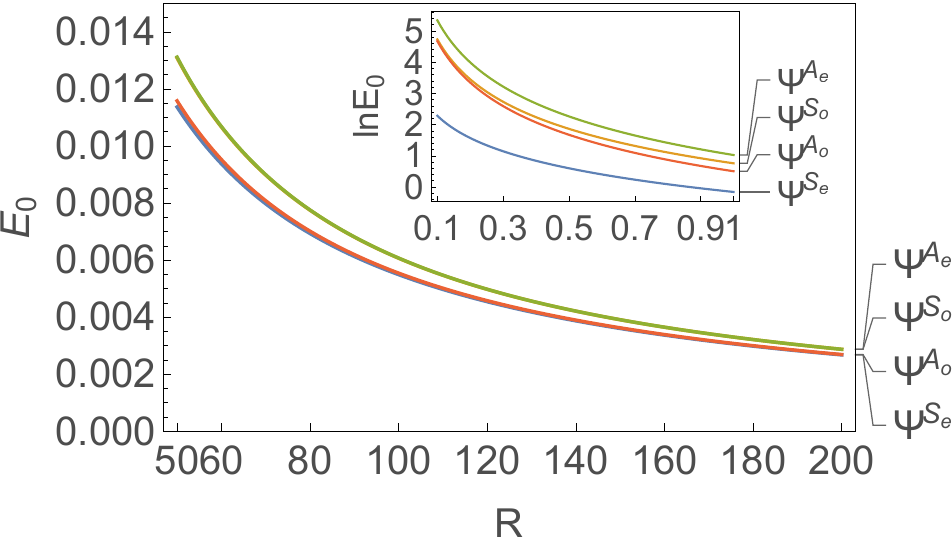}} 
\hspace{0.4in}
\subfigure[]{
  \includegraphics[width=0.35\textwidth]{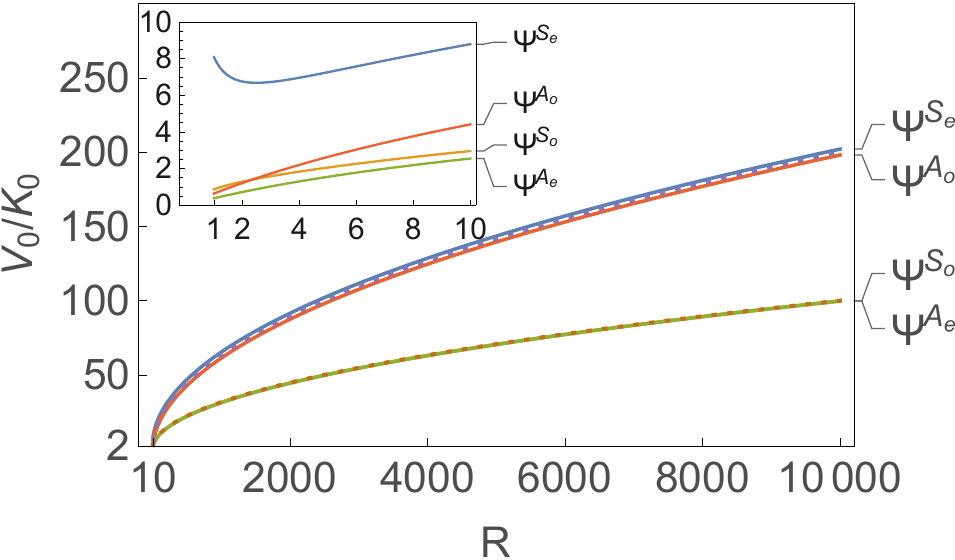} 
 }
\caption{Energetics analysis of the two-electron ground states of distinct symmetries. In
the notation $\Psi^{X_p}$ for the ground-state wave function, the superscript $X_p$ indicates that
it is exchange symmetric ($X=S$) or exchange antisymmetric
($X=A$), and has even ($p=e$) or odd ($p=o$) parity.
(a) Plot of the ground-state energy $E_0$ vs the radius $R$ of the sphere. Energy
degeneracies are found in the large- and small-$R$ regimes. (b) The potential energy $V_0$ dominates
over the kinetic energy $K_0$ in the large-$R$ regime. Solid lines are from the CI method. Dashed
lines are from the small-oscillation theory. }
\label{two_energy}
\end{figure}

\subsection{The case of two electrons}

For a two-electron system, the construction of the ground-state wave function must obey the
Pauli exclusion principle. The orbital wave function of the two-electron system is
either exchange-symmetric or exchange-antisymmetric depending on the spin state
of the electrons. We discuss both cases in this section.

\begin{figure*}
\centering
\subfigure[\
  $\Psi^{S_e}$]{\includegraphics[width=0.24\textwidth]{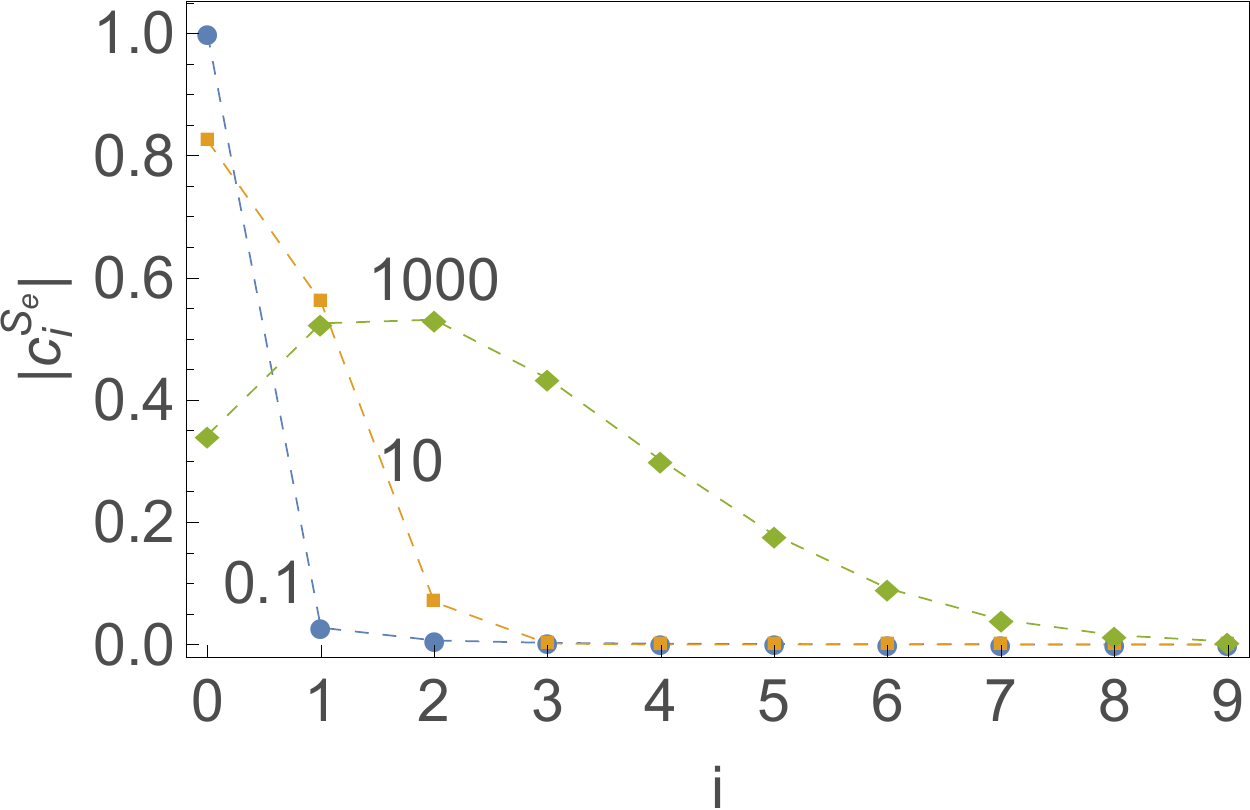}}
\subfigure[\
  $\Psi^{A_o}$]{\includegraphics[width=0.24\textwidth]{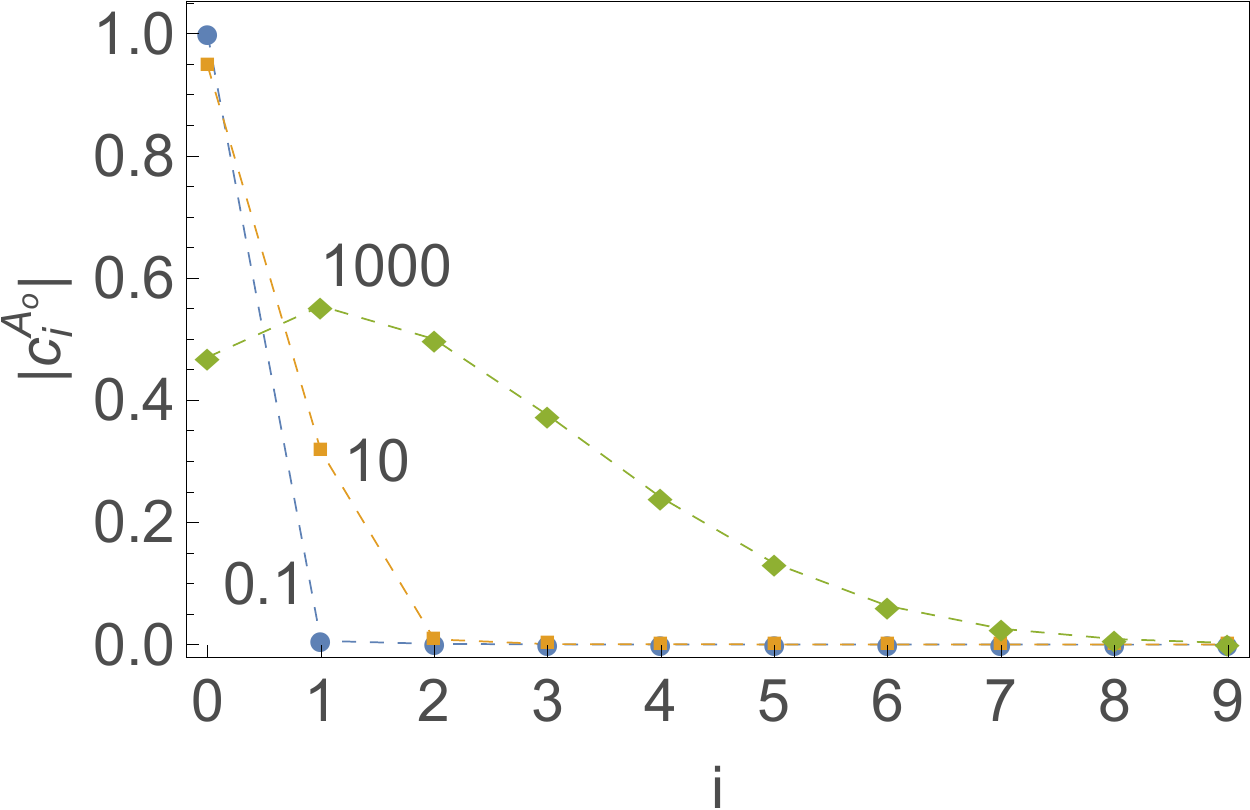}}
\subfigure[\
  $\Psi^{S_o}$]{\includegraphics[width=0.24\textwidth]{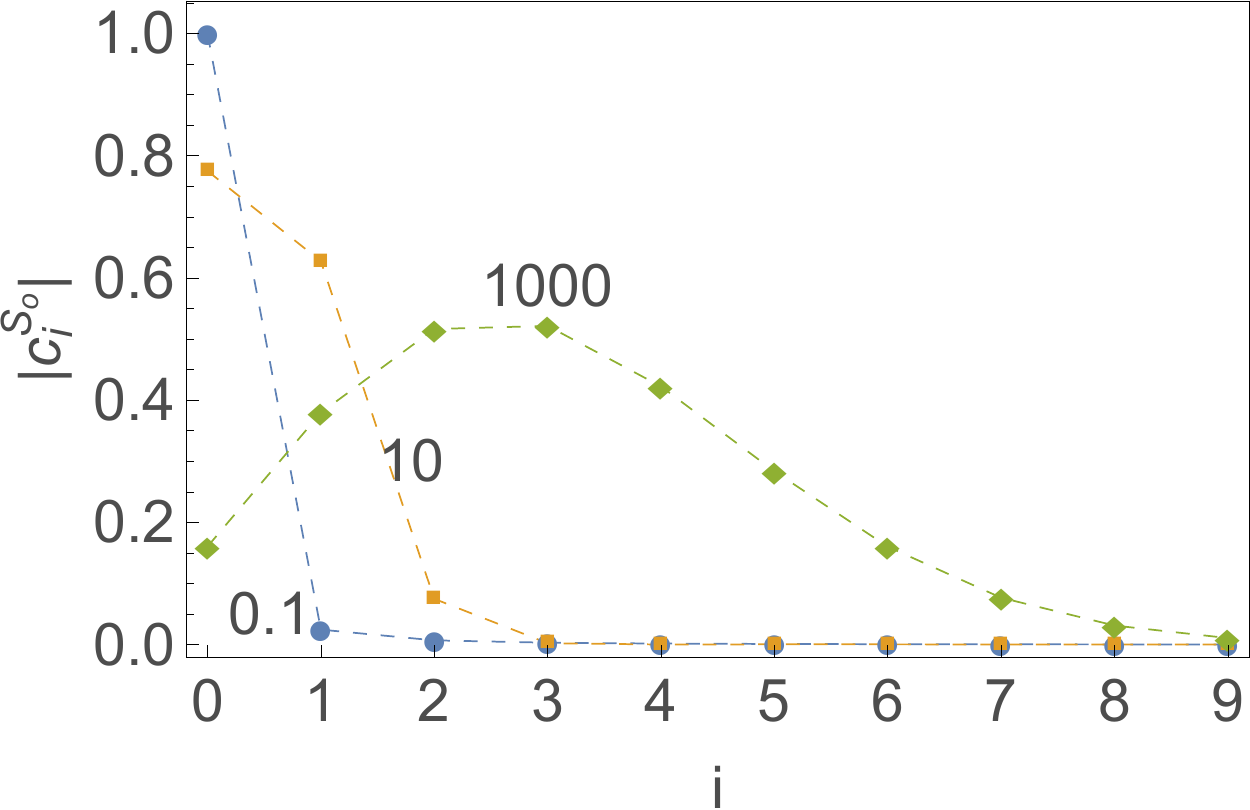}}
\subfigure[\
  $\Psi^{A_e}$]{\includegraphics[width=0.24\textwidth]{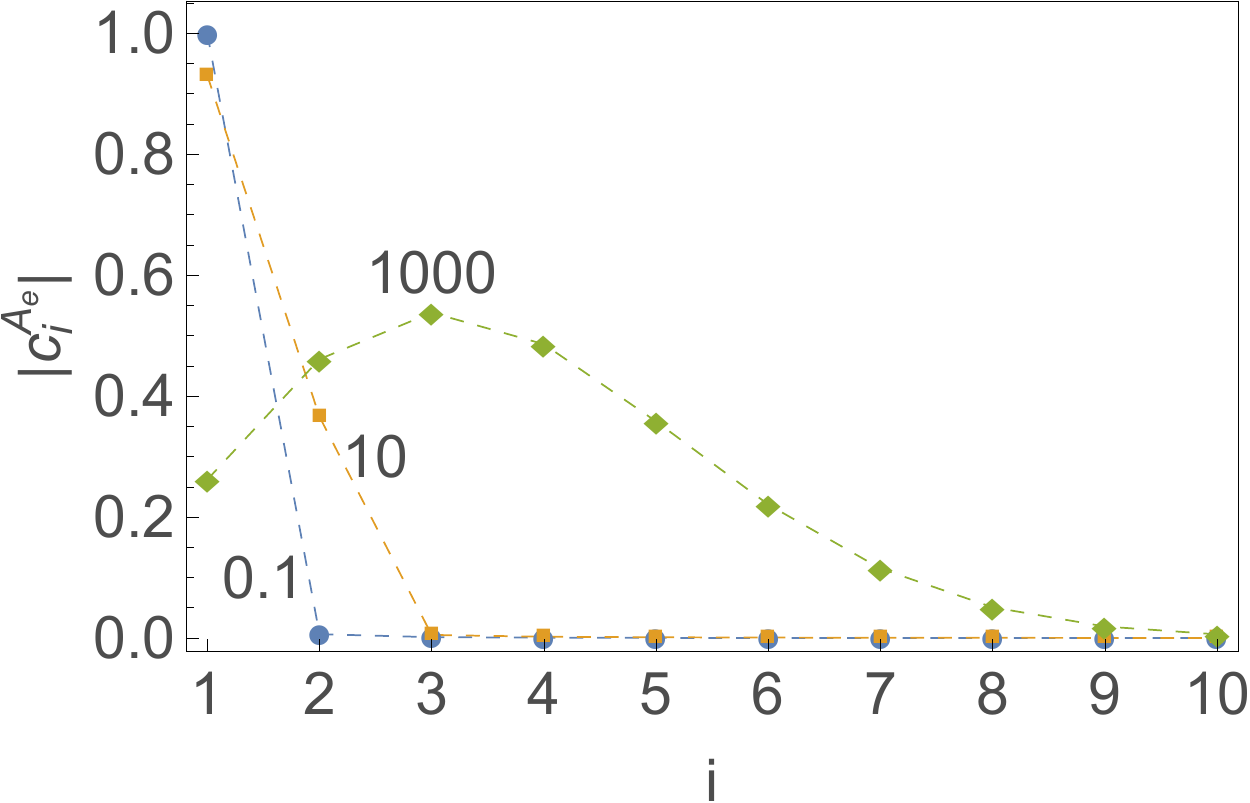}}
\subfigure[\
  $\Psi^{S_e}$]{\includegraphics[width=0.24\textwidth]{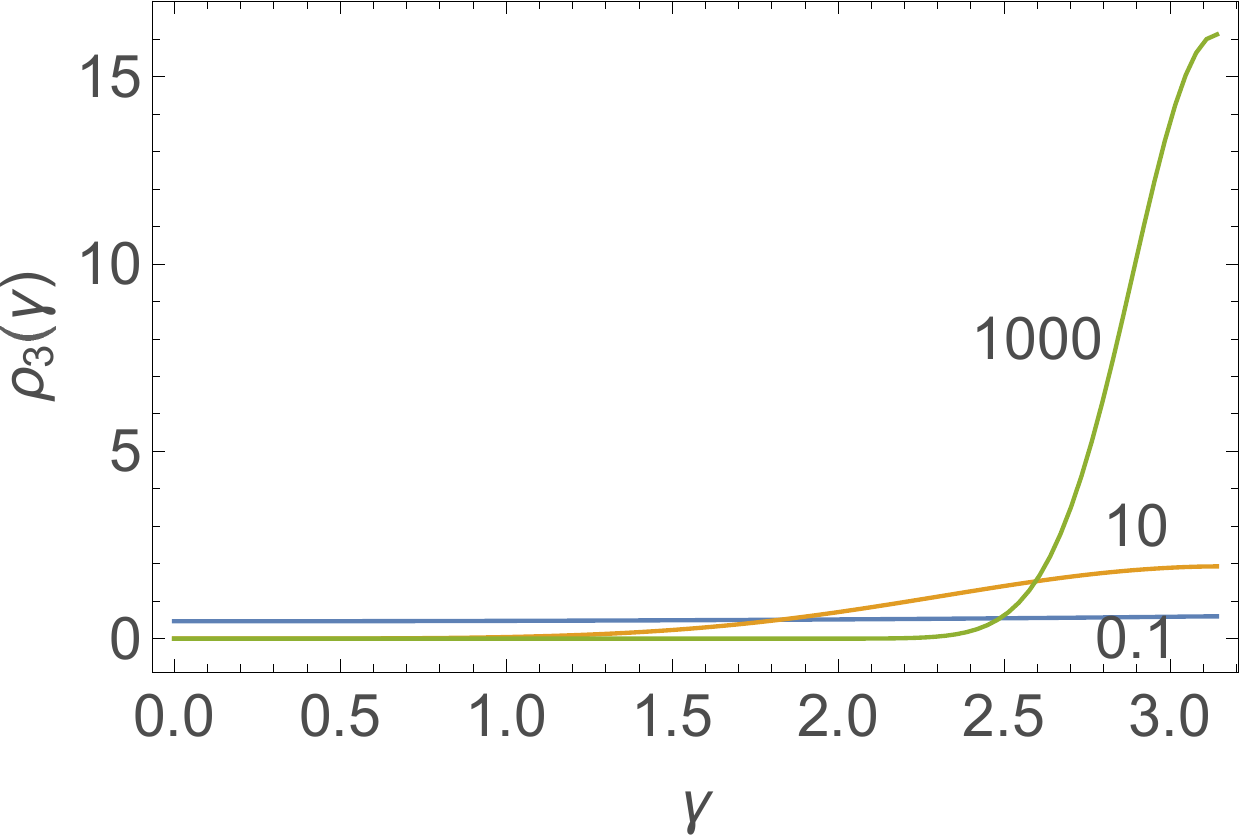}}
\subfigure[\
  $\Psi^{A_o}$]{\includegraphics[width=0.24\textwidth]{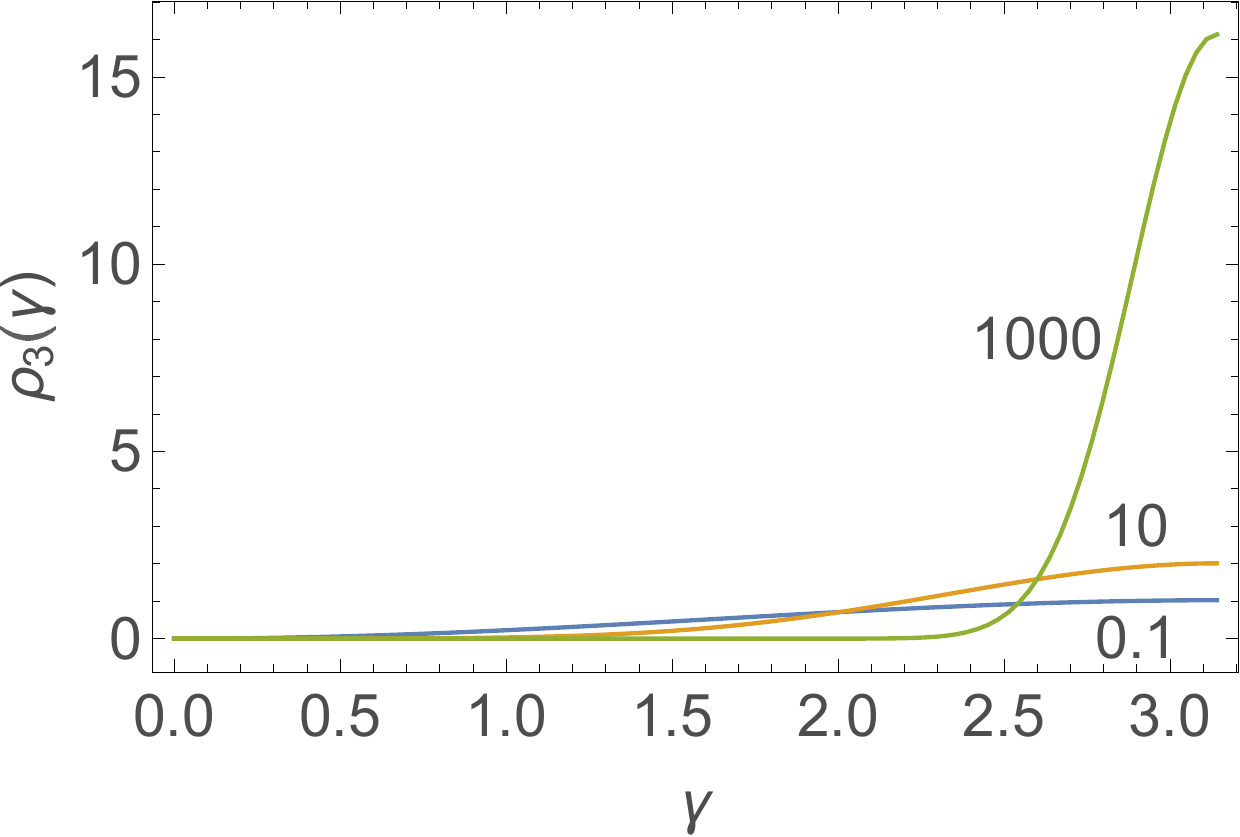}}
\subfigure[\
  $\Psi^{S_o}$]{\includegraphics[width=0.24\textwidth]{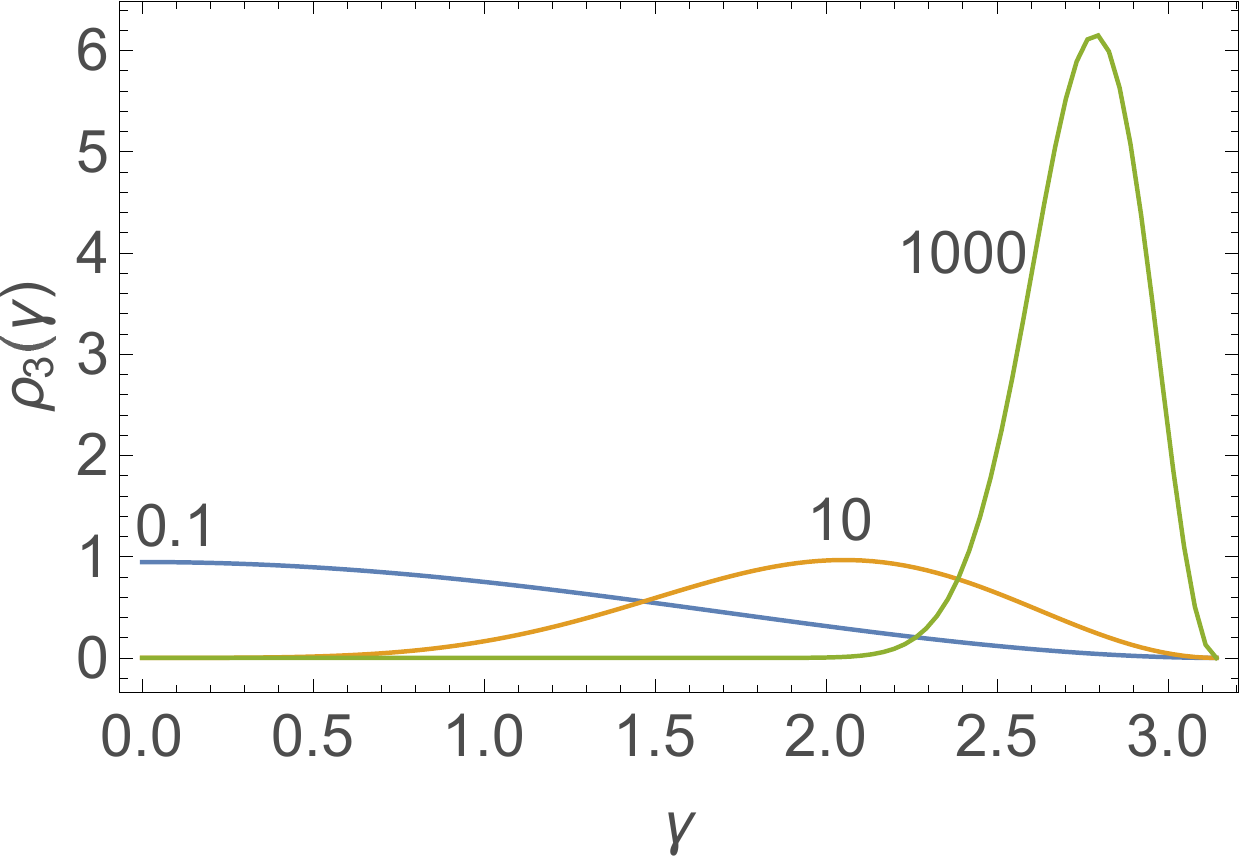}}
\subfigure[\
  $\Psi^{A_e}$]{\includegraphics[width=0.24\textwidth]{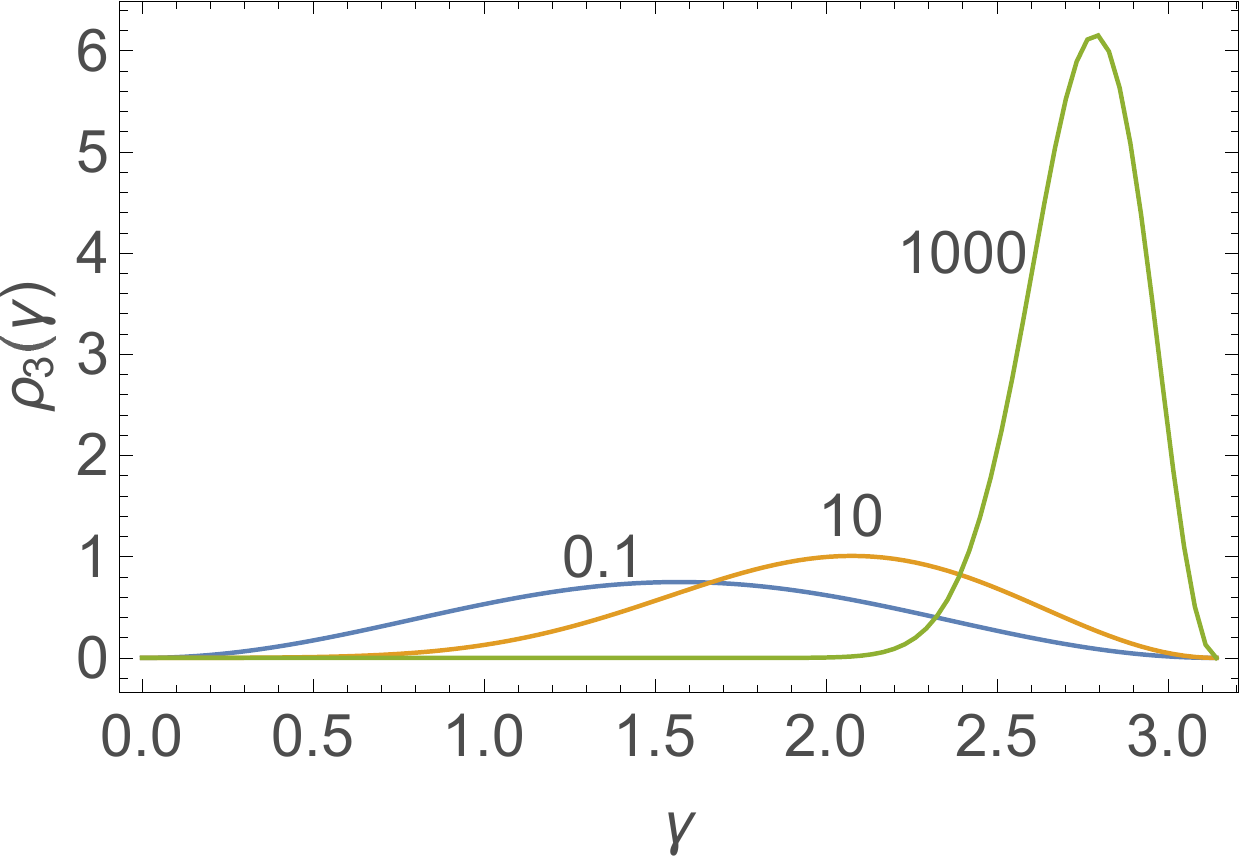}}
\caption{Analysis of the ground state wave functions for the two-electron system
constructed based on the CI method. The number near
each curve indicates the value of $R$. (a)-(d) Plot of the
amplitude of the angular momentum quantum number $i$ at varying $R$. (e)-(h) Distribution of the reduced
probability density $\rho_2(\gamma)$, which is defined in Eq.(\ref{defrho}). $\gamma$ is the angular distance of the two
electrons. }
\label{two_cross}
\end{figure*}

{\bf{Construction of symmetrized ground state wave functions}} For the two-electron
system, the common eigenstates $|l_1,l_2,l,m\rangle$ of $\hat{L}^2$, $\hat{L}_z$,
$\hat{L}_1^2$, $\hat{L}_2^2$ constitute the bases of the complete Hilbert space,
denoted as $\mathscr{H}$. Note that $|l_1,l_2,l,m\rangle$ can be constructed by direct products of single
particle states using Clebsch-Gordan coefficients~\cite{landau}. From the basis wave
function $\Psi_{\bold{n}}(\vec{r}_1,\vec{r}_2)$, where
$\bold{n}= (l_1,l_2,l,m)$, one can construct wave functions
$\Psi_{\bold{n}}^{X_p}(\vec{r}_1,\vec{r}_2)$ with a certain symmetry $X_p$.

To implement the CI method, we first notice that the Hilbert space $\mathscr{H}$ can be
reduced to the sum of the subspaces $\mathscr{H}(l,m)$:
$\mathscr{H}=~\mathop{\oplus}\limits_{l,m}\mathscr{H}(l,m)$. The subspace
$\mathscr{H}(l,m)$ is spanned by the bases $\varepsilon(l, m)$:
$\varepsilon(l,m)=\big\{|l_1,l_2,l,m\rangle\big | l_1,l_2 = 0, 1, 2 ... \big\}$, where
$|l_1-l_2|\leq l\leq l_1+l_2$ and $-l\leq m\leq l$. We search for the ground state wave
functions in the subspaces of $\mathscr{H}(0,0)$ and $\mathscr{H}(1,0)$. According to
angular momentum algebra, all the basis wave functions in $\mathscr{H}(0,0)$ are
exchange-symmetric and with even parity~\cite{2esph}. That is,
$\varepsilon(0,0)=\varepsilon^{S_e}(0,0)=\{|i,i,0,0\rangle\big|i = 0, 1, 2...\}$. 
The bases of the subspace $\mathscr{H}(1,0)$ can be classified according to their parity
and exchange symmetry: $\varepsilon(1,0)=\varepsilon^{S_o}(1,0)\oplus\varepsilon^{A_e}(1,0)\oplus\varepsilon^{A_o}(1,0)$. Specifically, 
$\varepsilon^{S_o}(1,0)=\{\frac{1}{\sqrt{2}}(|i,i+1,1,0\rangle+|i+1,i,1,0\rangle)\big|i = 0, 1, 2...\}$, 
$\varepsilon^{A_e}(1,0)=\{|i,i,1,0\rangle\big|i= 1, 2...\}$, 
$\varepsilon^{A_o}(1,0)=\{\frac{1}{\sqrt{2}}(|i,i+1,1,0\rangle-|i+1,i,1,0\rangle)\big|i = 0, 1, 2...\}$.
Note that no base of the subspace $\mathscr{H}(1,0)$ is both exchange symmetric and with
even parity. In fact, any wave function in $\mathscr{H}(1,0)$ with even parity must be
exchange antisymmetric (see SI).

We denote the basis states in $\mathscr{H}^{S_e}(0,0)$,
$\mathscr{H}^{S_o}(1,0)$, $\mathscr{H}^{A_e}(1,0)$ and $\mathscr{H}^{A_o}(1,0)$, as
$|\Psi_i^{X_p}\rangle$, where $i$ completely determines the values of $l_1, l_2, l, m$ in
$\varepsilon(l, m)$, as shown in preceding discussion. Any state in these subspaces can be
expressed as a linear superposition of $|\Psi_i^{X_p}\rangle$:
$|\Psi^{X_p}\rangle =\sum\limits_{i=i_{0}}^{i_{max}} c^{X_p}_i|\Psi^{X_p}_i\rangle$. In
our numerical construction of the ground-state wave functions, $i_0=1$ for
$|\Psi^{A_e}\rangle$, and $i_0=0$ for $|\Psi^{S_e}\rangle$, $|\Psi^{S_o}\rangle$, and
$|\Psi^{A_o}\rangle$. Here, $i_{max}=100$. We obtain the values for the coefficients
$\{c_i^{X_p}\}$ by expanding the Hamiltonian in the Hilbert space
spanned by $|\Psi^{X_p}_i\rangle$ and solving for the equation
$\sum\limits_{i'}H^{X_p}_{ii'}c^{X_p}_{i'}=E c^{X_p}_i$.

{\bf{Analysis of ground states}}
In Fig.~\ref{two_energy}(a), we show the monotonous decrease of the ground-state energy $E_0$
with the radius $R$ of the sphere for the four kinds of wave functions with
distinct symmetries. Among these four cases, the $\Psi^{S_e}$ state, which is in
the Hilbert subspace of
$\mathscr{H}(0,0)$, has the lowest energy. Note that our numerically solved energy of the $\Psi^{S_e}$
state agrees well with the previously
reported exact values~\cite{2esph} (see SI for more information).

Figure~\ref{two_energy}(a) also shows the degeneracy of the $\Psi^{S_o}$ and
$\Psi^{A_o}$ states in the small-$R$ regime; and the degeneracy of the
$\Psi^{S_e}$ and $\Psi^{A_o}$ states in the large-$R$ regime. As
$R\to\infty$, all the four energy curves in Fig.~\ref{two_energy}(a) tend to converge
towards $1/(2R)$. This result is consistent with the following scaling argument. Since
the kinetic energy is inversely proportional to $1/R^2$ and the potential energy is
inversely proportional to $1/R$, the potential energy will dominate, and the total energy
will scale with $R$ in the form of $1/R$ in the large-$R$ limit.

To show the relative contributions of the potential
and kinetic energies to the total energy $E_0$, we plot the $V_0/K_0$ vs 
$R$ curve in Fig.~\ref{two_energy}(b). With the increase of $R$, the potential
energy will dominate over the kinetic energy. We also see the merging of the
$\Psi^{S_e}$ and $\Psi^{A_o}$ and $\Psi^{S_o}$ and
$\Psi^{A_e}$ curves in the large-$R$ regime. The $V_0/K_0$ curves in
the small-$R$ regime are shown in the inset. Since the $\Psi^{S_e}$
state is in the subspace $\mathscr{H}(0,0)$, whose zeroth-order wave function has zero
kinetic energy, the $\Psi^{S_e}$ curve is obviously above all the other three
states in the $\mathscr{H}(1,0)$ space.

We further show the
contribution of each $|\Psi^{X_p}_i\rangle$ component in
the constructed ground state $|\Psi^{X_p}\rangle$ by analyzing the
coefficients $|c^{X_p}_i|$. In
Figs.~\ref{two_cross}(a)-\ref{two_cross}(d), we present the
values of $|c^{X_p}_i|$ for the first ten angular momentum quantum numbers $i$. The four
kinds of wave functions exhibit uniform behavior with the
increase of $R$. The value of the dominant $i$, which is zero at $R=0.1$, increases with
$R$. Meanwhile, increasing $R$ widens the $|c^{X_p}_i|$ curves. The underlying
physics is as follows: $|\Psi^{X_p}_i\rangle$-components of larger $i$
are excited under the increasingly important Coulomb interaction with the expansion of
the sphere.

To characterize the
correlation of the two electrons on the sphere, we compute the reduced two-electron
probability
density $\rho_2(\gamma)$ as a function of their angular distance $\gamma$~\cite{correlation2}:
\begin{align}\label{defrho}
  \rho_2(\gamma)=\int\int
  P(\vec{r}_1,\vec{r}_2)\delta(\hat{\vec{r}}_1\cdot \hat{\vec{r}}_2 - \cos{\gamma})dS_1dS_2,
\end{align} where $P(\vec{r}_1,\vec{r}_2)$ is the probability density of finding electron
1 and electron 2 simultaneously at $\vec{r}_1$ and $\vec{r}_2$ on the sphere. According
to Born's statistical interpretation of quantum mechanics,
$P(\vec{r}_1,\vec{r}_2)=|\Psi^{X_p}(\vec{r}_1,\vec{r}_2)|^2$.
Since $\int\int P(\vec{r}_1,\vec{r}_2)dS_1dS_2=1$, the normalization condition
for $\rho_2(\gamma)$ is $\int_0^{\pi}\rho_2(\gamma)\sin{\gamma}d\gamma=1$.

In Figs.~\ref{two_cross}(e)-\ref{two_cross}(h), we plot the $\rho_2(\gamma)$ curves for
all the four kinds of $\Psi^{X_p}(\vec{r}_1,\vec{r}_2)$. We see that when $R$ is small, the
correlation between the two electrons is relatively weak. With the increase of $R$, sharp
peaks on the $\rho_2(\gamma)$ curves are developed at $\gamma=\pi$ [see
Figs.~\ref{two_cross}(e) and \ref{two_cross}(f)] or near  $\gamma=\pi$
[Figs.~\ref{two_cross}(g) and \ref{two_cross}(h)], which indicates the enhanced
electron-electron correlation. These two kinds of
electron localization at and near diametric poles correspond to different vibration
modes, as will be shown in the next section. A comparison of
Figs.~\ref{two_cross}(a)-\ref{two_cross}(d) and \ref{two_cross}(e)-\ref{two_cross}(h)
shows that the localization of the electrons at the diametric poles accompanies the
widening of the $|c^{X_p}_i|$ curves. In other words, a strongly correlated electron
state results from a combination of multiple monochromatic states.  The configuration of
two highly localized diametric electrons found on a large sphere is
consistent with the preceding energetics analysis, and it has connections to Wigner
crystallization occurring in the two-dimensional electron gas in a uniform, neutralizing
background when the electron density is less than a critical value~\cite{crystal}.

The correlation between the two electrons can also be characterized by the 
mean inverse separation $\tilde{d}_{12}^{-1}$, which is defined as
$\tilde{d}_{12}^{-1}=\int\int\frac{R}{|\vec{r}_1-\vec{r}_2|}|\Psi(\vec{r}_1,\vec{r}_2)|^2dS_1dS_2$.
It is recognized that $\tilde{d}_{12}^{-1}=V_0R$. For all four kinds of
$\Psi^{X_p}(\vec{r}_1,\vec{r}_2)$, we numerically show that $\tilde{d}_{12}^{-1}$
decreases monotonously with $R$, and asymptotically to $1/2$ in the large-$R$ limit.

{\bf{Asymptotic behaviors in the small- and large-$R$ regimes}} 
In this section, we perform perturbation analysis in the small-$R$ regime, and propose
small oscillation theory (which is also called ``strong-coupling perturbation
theory"~\cite{seidl2007adiabatic}) in the large-$R$ regime to discuss the asymptotic
behaviors of the two-electron system.  The presented theoretical results can also be used
to rationalize the energy curves in Fig.~\ref{two_energy}.

We first apply perturbation theory to analyze
the ratio of the potential and kinetic energies $V_0/K_0$ for all four kinds of wave
functions $\Psi^{X_p}(\vec{r}_1,\vec{r}_2)$ in the small-$R$ regime. The total angular
momentum quantum number in the unperturbed state (Coulomb
interaction is turned off) is denoted $i_0$. The ground-state energy and wave function can
be written as
\begin{align}
E_0=E_0^{(0)}+E_0^{(1)}+E_0^{(2)}+...
\end{align}
\begin{align}
|\Psi^{X_p}\rangle=|\Psi^{X_p(0)}\rangle+|\Psi^{X_p(1)}\rangle+|\Psi^{X_p(2)}\rangle+...
\end{align}
where $E^{(0)}_0=K^{X_p}_{i_0i_0}$, and $E^{(1)}_0=V^{X_p}_{i_0i_0}$.
$\Psi^{X_p(1)}(\vec{r}_1,\vec{r}_2) = \sum_{i>i_0} c^{(1)}_i
\Psi^{X_p}_i(\vec{r}_1,\vec{r}_2)$, where
$c^{(1)}_i=V^{X_p}_{i_0i}/[i_0(i_0+1)-i(i+1)]$.

Keeping up to the first-order term, we have 
\begin{align}\label{small}
  \frac{V_0}{K_0}
  =\frac{R^2V_{i_0i_0}^{X_p}}{i_0(i_0+1)+\sum\limits_{i>i_0}i(i+1)|c^{(1)}_i|^2}\nonumber\\
\end{align}
Note that both $RV_{i_0i_0}^{X_p}$ and $c^{(1)}_i/R$ are independent of $R$. Here, $i_0=1$ for
$\Psi^{A_e}(\vec{r}_1,\vec{r}_2)$, and $i_0=0$ for $\Psi^{S_e}(\vec{r}_1,\vec{r}_2)$. For $\Psi^{S_e}(\vec{r}_1,\vec{r}_2)$,
$V_0/K_0 \sim 1/R$. For
$\Psi^{S_o}(\vec{r}_1,\vec{r}_2)$, $\Psi^{A_e}(\vec{r}_1,\vec{r}_2)$, and $\Psi^{A_o}(\vec{r}_1,\vec{r}_2)$,
$V_0/K_0\sim R$. These scaling laws are consistent with the inset in
Fig. \ref{two_energy}(b).

On a large sphere, the localized electrons at
diametric poles, as shown in Figs.~\ref{two_cross}(e)-\ref{two_cross}(h), are
inevitably subject to small vibration due to Heisenberg's uncertainty principle. 
The two-particle case has been analyzed in the quantum regime~\cite{2el}.
However, it is a challenge to generalize the quantum treatment to
multiple-particle cases. Here, we perform a semi classical analysis of the small
vibration of the electrons that can be readily extended to the three-electron
case.

The classical Hamiltonian to describe the small vibration of two particles around the
diametric equilibrium positions at $(\bar{\theta}_1=\pi/2,\bar{\phi}_1=0)$ and
$(\bar{\theta}_2=\pi/2,\bar{\phi}_2=\pi)$ on the sphere is 
\begin{eqnarray}
  H=\frac{1}{2} R^2 \sum\limits_{i=1}^2(\delta\dot{\theta}_i^2 + \delta\dot{\phi}_i^2)
   +\sum\limits_{i,j=1}^2(\delta\theta_i\delta\theta_jD^1_{ij}&+&\delta\phi_i\delta\phi_jD^2_{ij})\nonumber
   \\
   &+&\frac{1}{2R}.
  \label{two_H_1}
\end{eqnarray}
where $D^1=\frac{1}{8R}\left(                 
  \begin{array}{cc}   
1 &  1\\  
  1   &1\\  
  \end{array}
\right)                 
$,
$  
D^2=\frac{1}{8R}
\left(                 
  \begin{array}{cc}   
1&-1\\  
-1&1\\  
  \end{array}
\right)                
$. 

With the orthogonal transformation
\begin{align}
\vartheta_1&=\frac{1}{\sqrt{2}}(\delta\theta_1+\delta\theta_2)\quad\vartheta_2=\frac{1}{\sqrt{2}}(\delta\phi_2-\delta\phi_1)\nonumber\\
\vartheta_3&=\frac{1}{\sqrt{2}}(\delta\theta_1-\delta\theta_2)\quad\vartheta_4=\frac{1}{\sqrt{2}}(\delta\phi_2+\delta\phi_1),\nonumber
\end{align}
Eq.(\ref{two_H_1}) becomes
\begin{eqnarray}
  H=  \frac{1}{2}R^2\sum\limits_{r=1}^4\dot{\vartheta}_r^2
  +\frac{1}{2}\sum\limits_{r=1}^2\omega^2R^2\vartheta_r^2 + \frac{1}{2R},
  \label{two_H_2}
\end{eqnarray}
where $\omega=R^{-\frac{3}{2}}/2$. $\vartheta_1$ and $\vartheta_2$ describe
the relative vibration of the two electrons. The
vibrational energy [the second term in Eq.(\ref{two_H_2})] is proportional to
$R^{-\frac{3}{2}}$. $\vartheta_3$ and $\vartheta_4$ describe the
rotation of the whole system around the $y-$ and $z$-axes,
respectively. The rotational
energy is proportional to $L(L+1)/R^2$, where $L$ is the angular momentum. The energy contribution from the 
rotation of the whole system [i.e., the terms associated with $\vartheta_3$ and
$\vartheta_4$ in the first sum term of Eq. (\ref{two_H_2}) can be ignored in comparison with that from the vibration of
the electrons. We focus on the relative vibration of the
electrons in the following discussion.

Quantization of the reduced Hamiltonian in Eq. (\ref{two_H_2}) leads to the expression
for the energy level of the
two-electron system in the large-$R$ regime~\cite{landau}:
\begin{equation} 
  E_{\{n\}}=(n+1)\hbar \omega + \frac{1}{2R}.\label{two_H_3}
\end{equation}
From Eq.(\ref{two_H_3}), by the virial theorem, we obtain the asymptotic expression for
${V_0}/{K_0}$ at large $R$: ${V_0}/{K_0} \sim R^{\frac{1}{2}}$. Our numerical results
based on the CI method conform to this scaling law, as shown in Fig.~\ref{two_energy}(b).

The value of the quantum number $n$ in Eq. (\ref{two_H_3}) can be obtained from the number of peaks in the
$\rho_2(\gamma)$ curve in Figs.~\ref{two_cross}(e)-\ref{two_cross}(h). For the cases of
$\Psi^{S_e}$ and $\Psi^{A_o}$ in Figs.~\ref{two_cross}(e) and
\ref{two_cross}(h) (see the curves for $R=1000$), the most probable positions of the
electrons are at the diametric poles, which correspond to the state of $n=0$. In
contrast, $n=1$ for the other two systems, as shown in Figs.~\ref{two_cross}(f) and
\ref{two_cross}(g) (see the curves for $R=1000$), where the most probable angular distance
of the two electrons slightly deviates from $\pi$.  Therefore, by their vibration
modes, the four kinds of
states $\Psi^{X_p}$ can be classified into two categories with $n=0$ and $n=1$. This
conclusion derived from small-oscillation theory is consistent with the result of
perturbative analysis of Schr\"{o}dinger's equation in the large-$R$ regime~\cite{2el}. The classification of the
states $\Psi^{X_p}$ by the vibration modes can account for the degeneracies
between $\Psi^{S_e}$ and $\Psi^{A_o}$ and $\Psi^{S_o}$ and
$\Psi^{A_e}$ in the large-$R$ regime, as shown in Fig.~\ref{two_energy}(a).

\begin{figure}
\centering
\subfigure[]
{\includegraphics[width=0.3\textwidth]{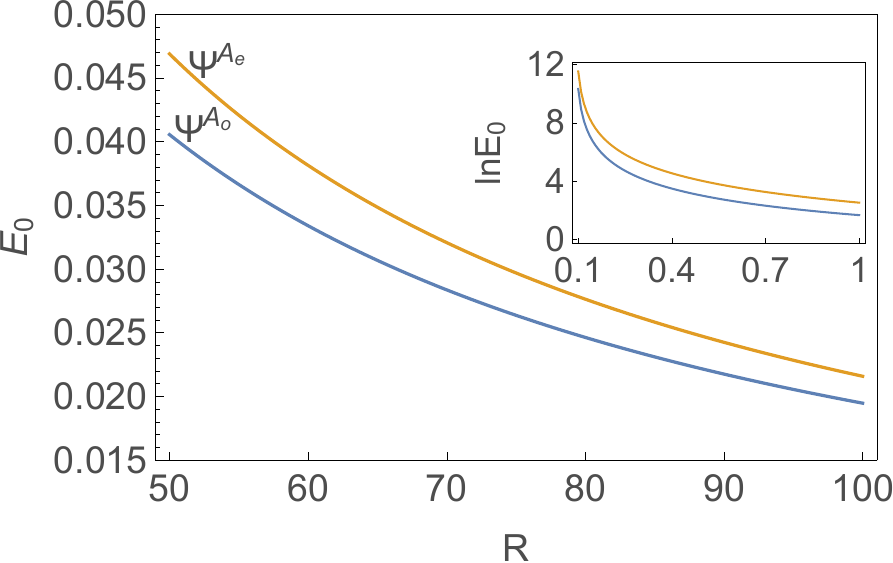}}
\hspace{0.4in}
\subfigure[]{
  \includegraphics[width=0.3\textwidth]{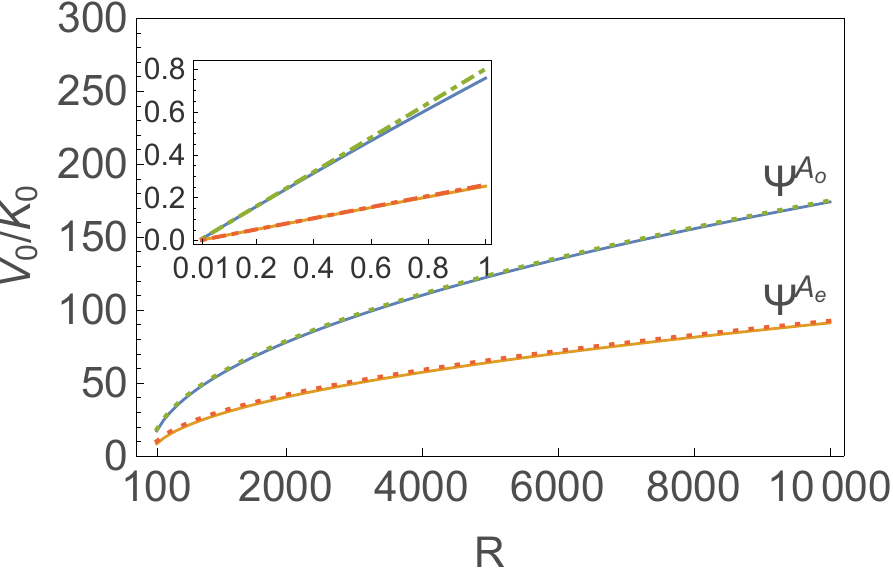} 
 }
\caption{ Energetics analysis of the three-electron ground states of distinct symmetries.
(a) Plot of the ground-state energy $E_0$ vs the radius $R$ of the sphere.
  (b) The potential energy $V_0$ dominates over the kinetic energy $K_0$ in the large-$R$ regime.
  Solid lines are from the CI method. Dashed lines conform to a power law of exponent
  $3/2$, as derived from the small-oscillation theory. Dot-dashed lines (in inset) are
  linear functions of $R$ according to perturbation theory for small $R$. }
\label{three_energy}
\end{figure}

\begin{figure}[t]
\centering
\subfigure[\ $\Psi^{A_o}$]{\includegraphics[width=0.3\textwidth]{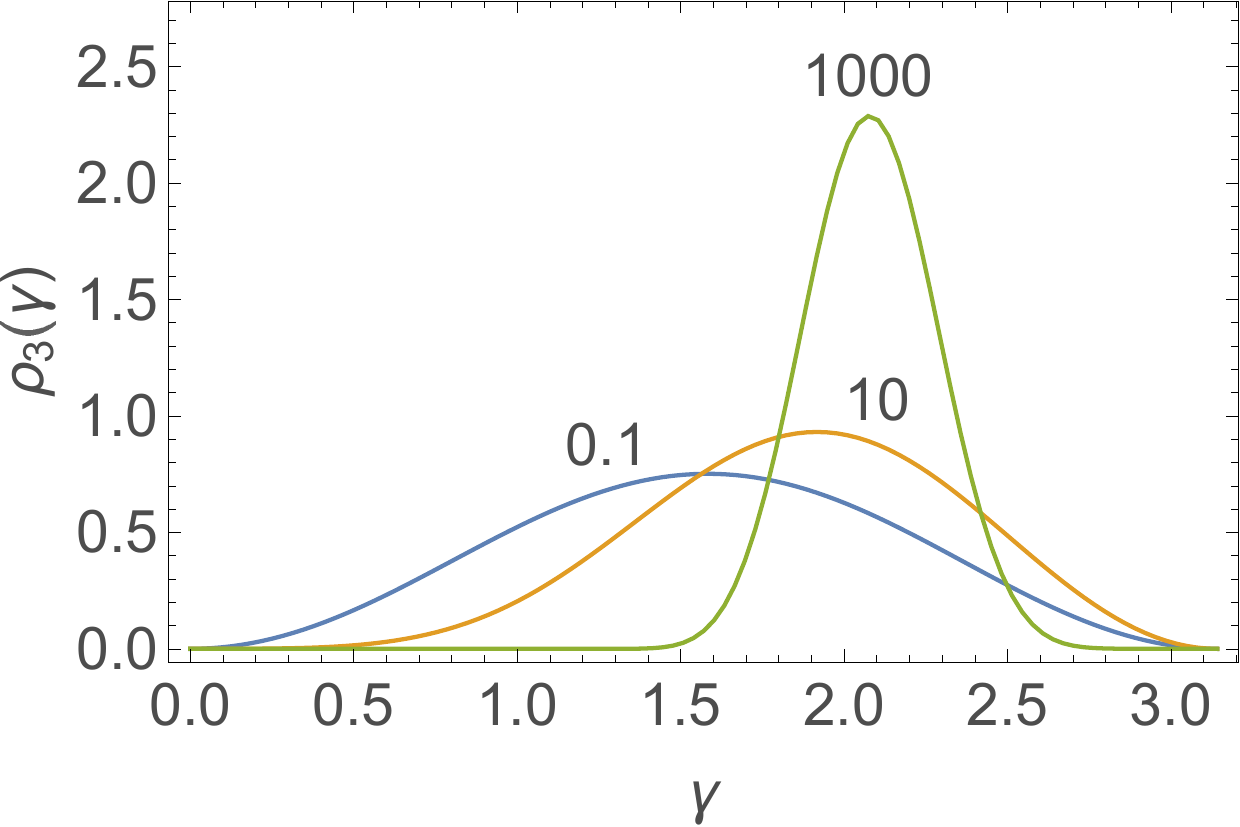}}
\subfigure[\ $\Psi^{A_e}$]{\includegraphics[width=0.3\textwidth]{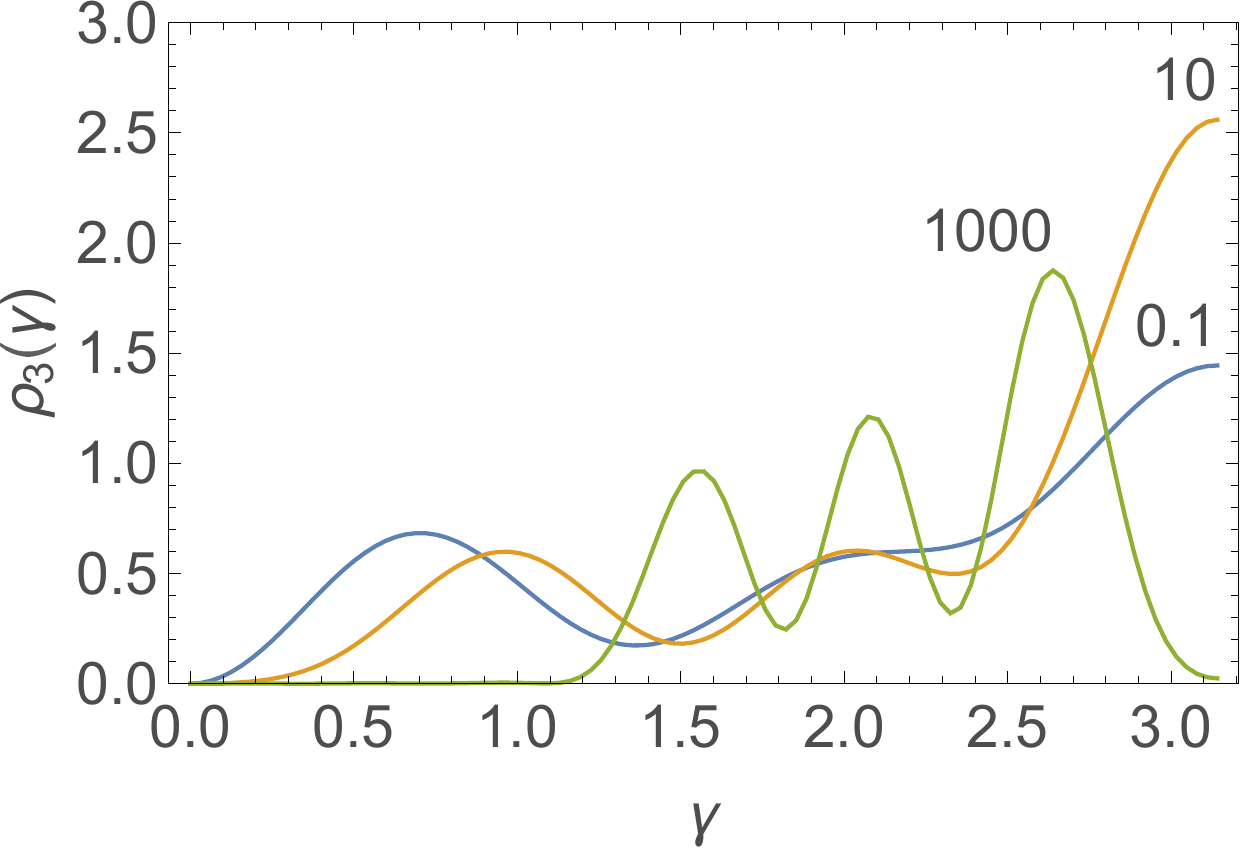}}
\caption{Distribution of the reduced probability density $\rho_3(\gamma)$ in the
three-electron system. $\rho_3(\gamma)$ is the probability of finding any two of the three
electrons with angular separation $\gamma$. 
  }\label{three_rho}
\end{figure}

\subsection{The case of three electrons}

For the three-electron system, we focus on the case of identical spin states. The ground-state wave functions must be exchange antisymmetric. We consider both odd and
even parities. These ground-state wave functions are denoted as $\Psi^{A_p}$, where $p=e$
(even parity) or $p=o$ (odd parity). In comparison with the two-electron system, we find
richer vibration modes for the three-electron system in the large-$R$ regime.

{\bf{Construction of symmetrized ground-state wave functions}}
For the three-electron system, $\hat{L}_{12}^2$, $\hat{L}_{1}^2$, $\hat{L}_{2}^2$,
$\hat{L}_{3}^2$, $\hat{L}^2$, and $\hat{L}_z$ commute with each other. Their common eigenstate $|l_{12},l_1,l_2,l_3,l,m\rangle$ is denoted as $|\mathbf{n}\rangle$. These eigenstates constitute the
basis set $\varepsilon(l,m)$ of
the Hilbert space $\mathscr{H}$.
$\mathscr{H}=\mathop{\oplus}\limits_{l,m}\mathscr{H}(l,m)$.
$\varepsilon(l,m)=\big\{|l_{12},l_1,l_2,l_3,l,m\rangle\big|l_1,l_2,l_3,l_{12} = 0, 1,
2...\big\}$, where $|l_1-l_2|\leq l_{12}\leq l_1+l_2, |l_{12}-l_3|\leq l\leq l_{12}+l_3,
-l\leq m\leq l$ (see SI). We construct the ground-state wave functions in the subspace
$\mathscr{H}(0, 0)$. $\varepsilon(0,0) = \{ |l_{12} = l_3, l_1,l_2,l_3, l=0, m =0 \rangle \}$,
where the
first equality is due to the zero total angular momenta. The basis set
$\varepsilon(0,0)$ is completely determined by $\bold{L} = (l_1, l_2, l_3)$.

By the standard coupling
of three angular momentums, we obtain the eigenstate wave
function of the three-electron system (see SI):
\begin{eqnarray}\label{3bases}
\Psi_{\bold{L}}(\vec{r}_1,\vec{r}_2,\vec{r}_3)&=
\sum\limits_{m_1=-l_1}^{l_1}\sum\limits_{m_2=-l_2}^{l_2}\left(\begin{array}{ccc}l_1&
l_2 & l_3 \\ m_1 & m_2 & -m_1-m_2\end{array}\right) \nonumber\\
&\times\frac{1}{R^3}Y_{l_1}^{m_1}(\vec{n}_1)Y_{l_2}^{m_2}(\vec{n}_2)Y_{l_3}^{-m_1-m_2}(\vec{n}_3).
\end{eqnarray}
Note that by Eq. (\ref{3bases}), $\Psi_{l_1l_2l_3}(\hat{P}_{ijk}\{\vec{r}_1,\vec{r}_2,\vec{r}_3\})=(-1)^{s(l_1+l_2+l_3)}\Psi_{\hat{P}^{-1}_{ijk}\{l_1l_2l_3\}}(\vec{r}_1,\vec{r}_2,\vec{r}_3)$. $\hat{P}_{ijk}$ is a permutation operator that changes the subscripts 1, 2, 3 of $r_m$ to $i$, $j$, $k$, respectively. $\hat{P}^{-1}_{ijk}$ changes the subscripts $i$, $j$, $k$ of $\ell_m$ to 1, 2, 3, respectively. $s=0$ and $1$ for even and odd permutations, respectively. This equation indicates that the new wave function is still in
the basis set $\varepsilon(0,0)$ under the permutation of the three electrons. In the
numerical construction of symmetrized ground
state wave functions, we use 1360 allowed odd parity bases from $\Psi^{A_o}_{l_1, l_2,
l_3}$ ($l_1 \in [1, 29]$, $l_2, l_3 \in [1, 30]$) and 1120 allowed even parity bases from
$\Psi^{A_e}_{1,2,3}$ ($l_1 \in [1, 27]$, $l_2, \in [2, 29]$, $l_3 \in [3, 30]$).

{\bf{Analysis of ground states}}
In Fig.~\ref{three_energy}(a), we show the monotonous decrease of the ground state energy
$E_0$ with $R$ for both cases of $\Psi^{A_o}$ and
$\Psi^{A_e}$. In the large-$R$ limit, both curves tend to converge
towards $\sqrt{3}/R$, which is the potential energy of three classical electrons sitting on the
vertices of a regular triangle circumscribed by the equator. In this asymptotic process
up to $R=10000$,
the ground-state energy of the $\Psi^{A_o}$ state is always slightly lower
than that of the $\Psi^{A_e}$ state. 

Figure~\ref{three_energy}(b) shows the ratio of the potential and kinetic
energies for both kinds of wave functions. Similar to the case
of the two-electron system, the potential energy will dominate over
the kinetic energy with $R$, suggesting the Coulomb-potential-driven
localization of electrons in the large-$R$ limit.

\begin{figure*}
\centering
\subfigure[\ $\Psi^{A_o}$, $R=1$]
{ \includegraphics[width=0.22\textwidth]{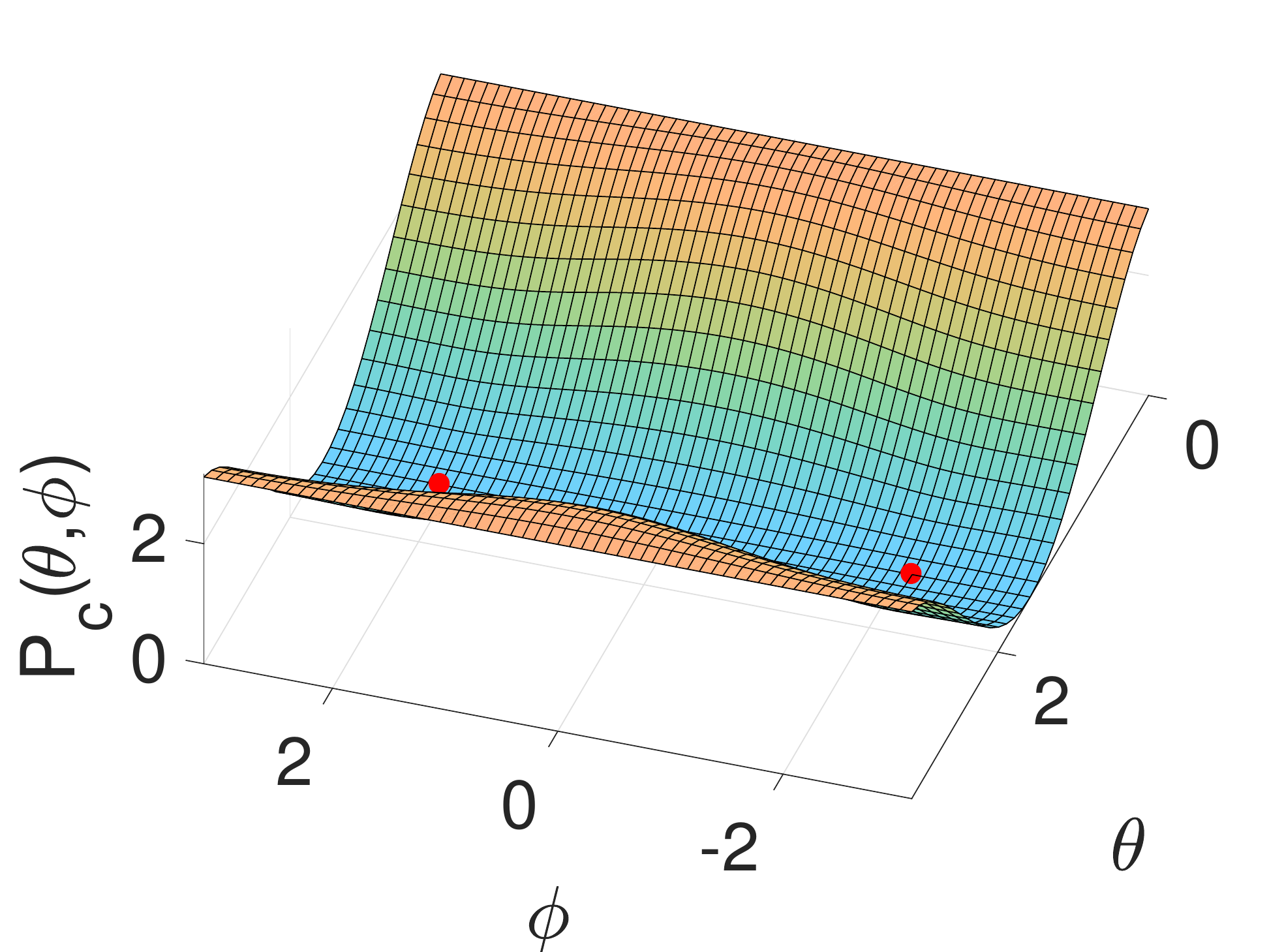}}\hspace{0.1 in}
\subfigure[\ $\Psi^{A_o}$, $R=10$]
{ \includegraphics[width=0.22\textwidth]{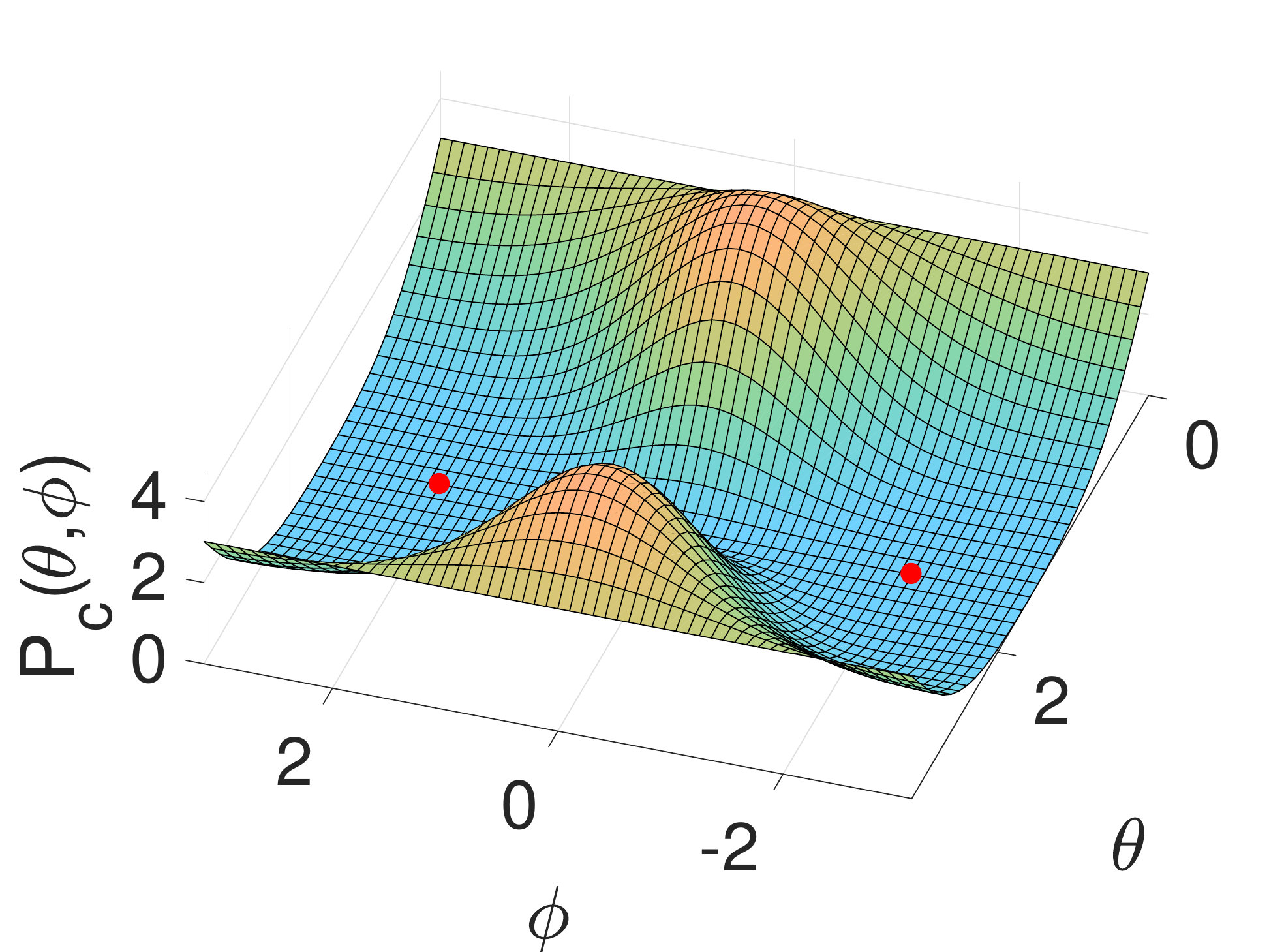}}\hspace{0.1 in}
\subfigure[\ $\Psi^{A_o}$, $R=100$]{ \includegraphics[width=0.22\textwidth]{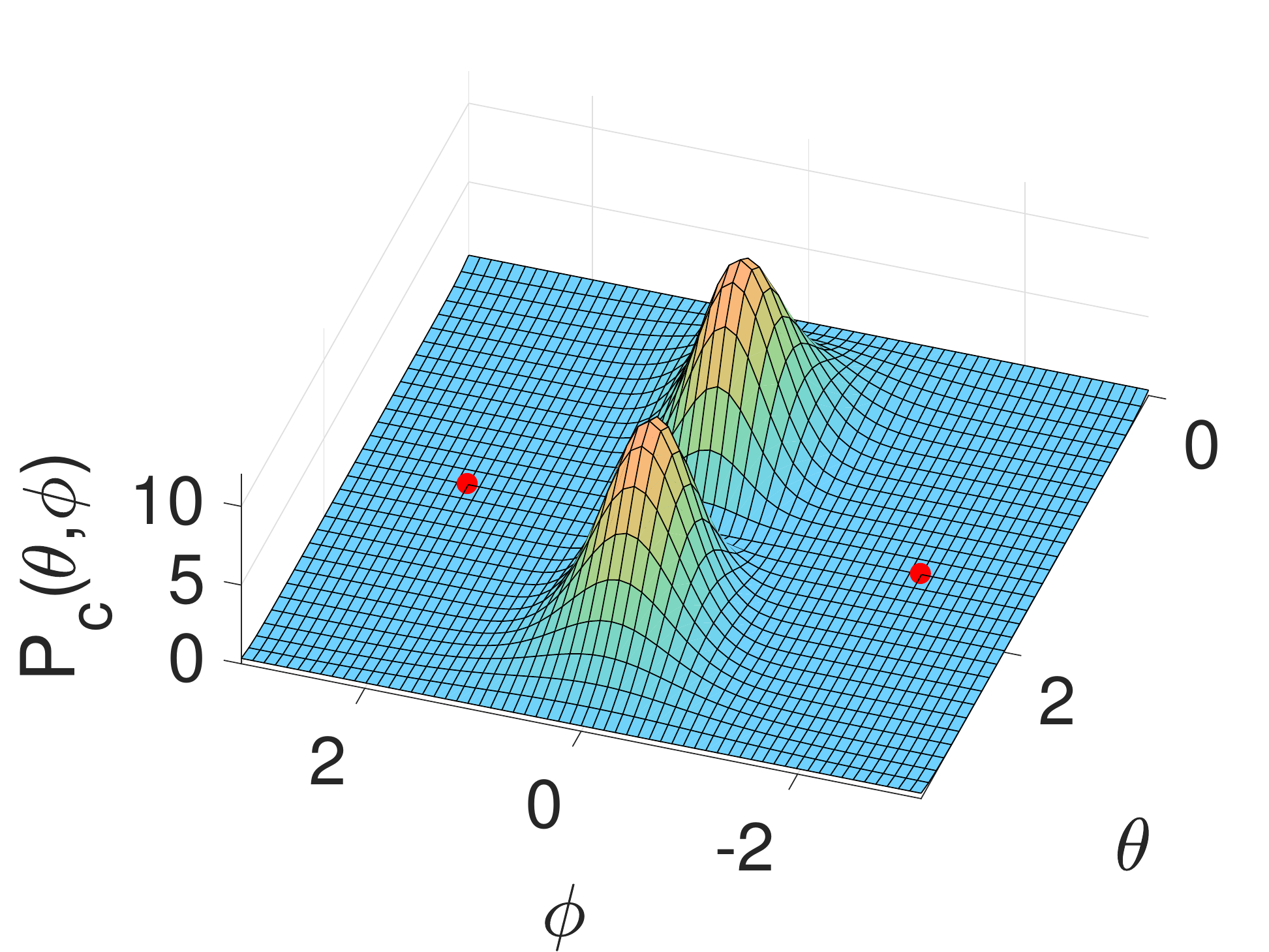}} \hspace{0.1 in}
\subfigure[\ $\Psi^{A_o}$, $R=1000$]{ \includegraphics[width=0.22\textwidth]{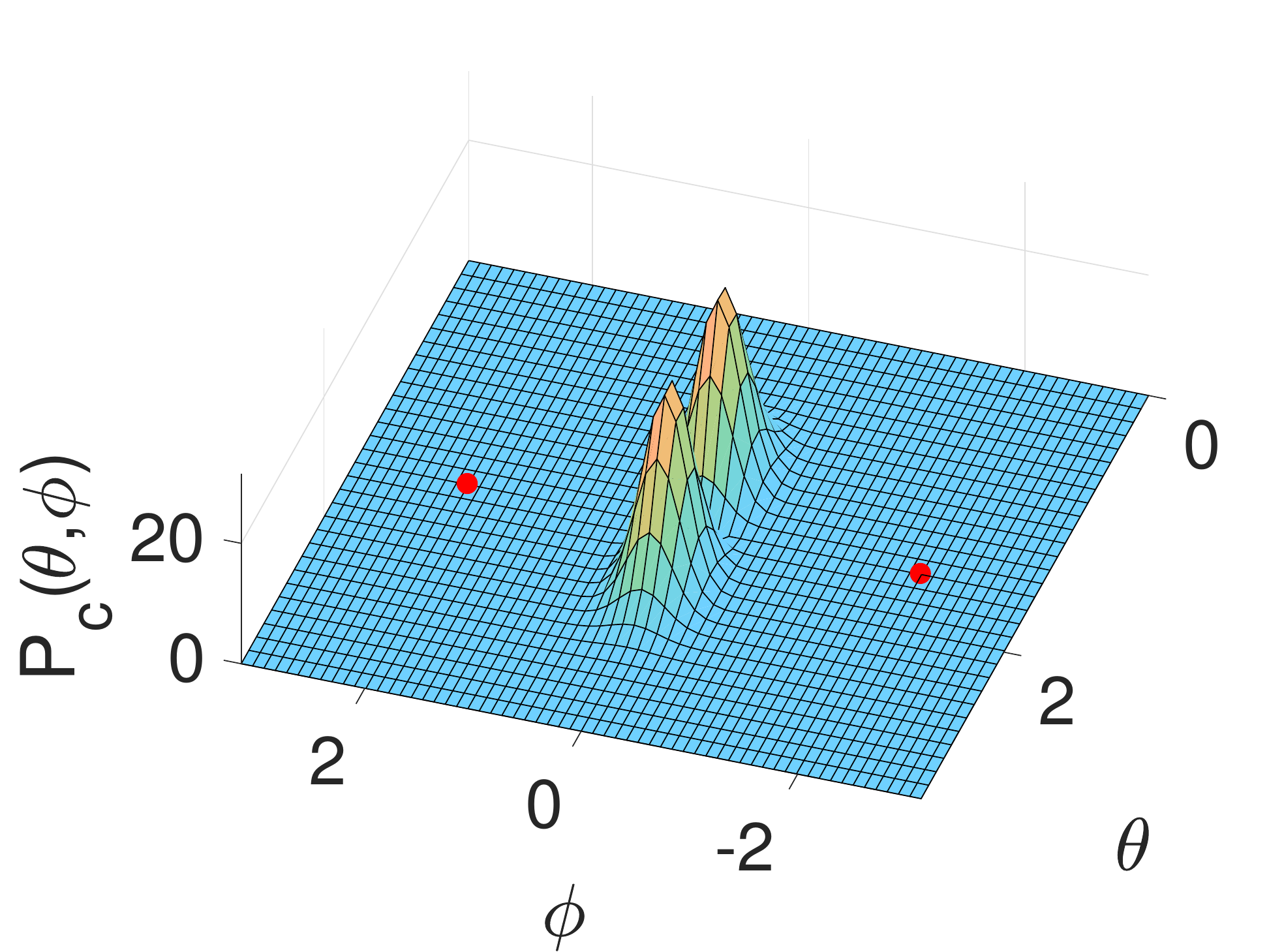}} \hspace{0.1 in}

  \subfigure[\ $\Psi^{A_e}$, $R=1$]
{ \includegraphics[width=0.22\textwidth]{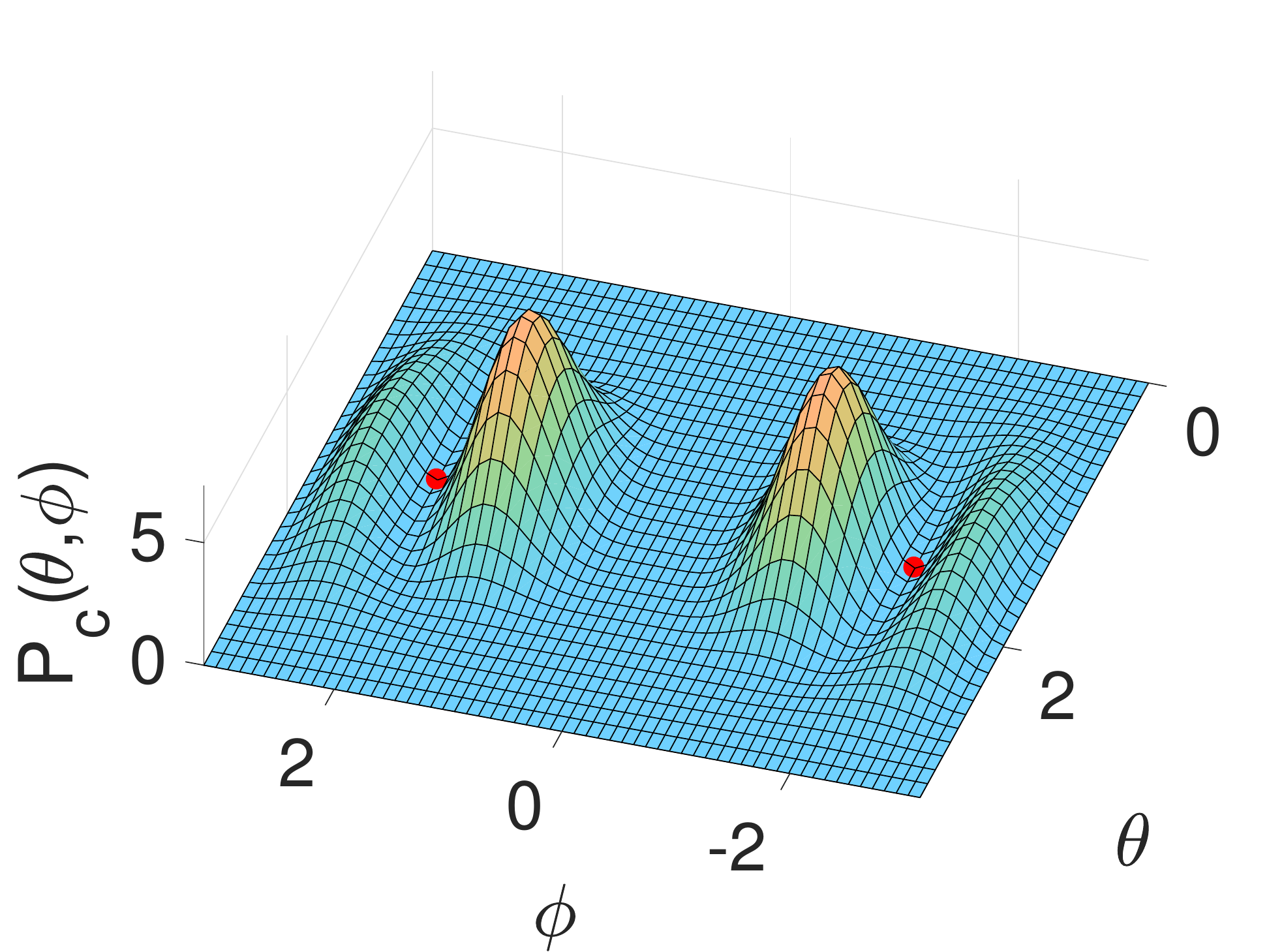}}\hspace{0.1 in}
\subfigure[\ $\Psi^{A_e}$, $R=10$]
{ \includegraphics[width=0.22\textwidth]{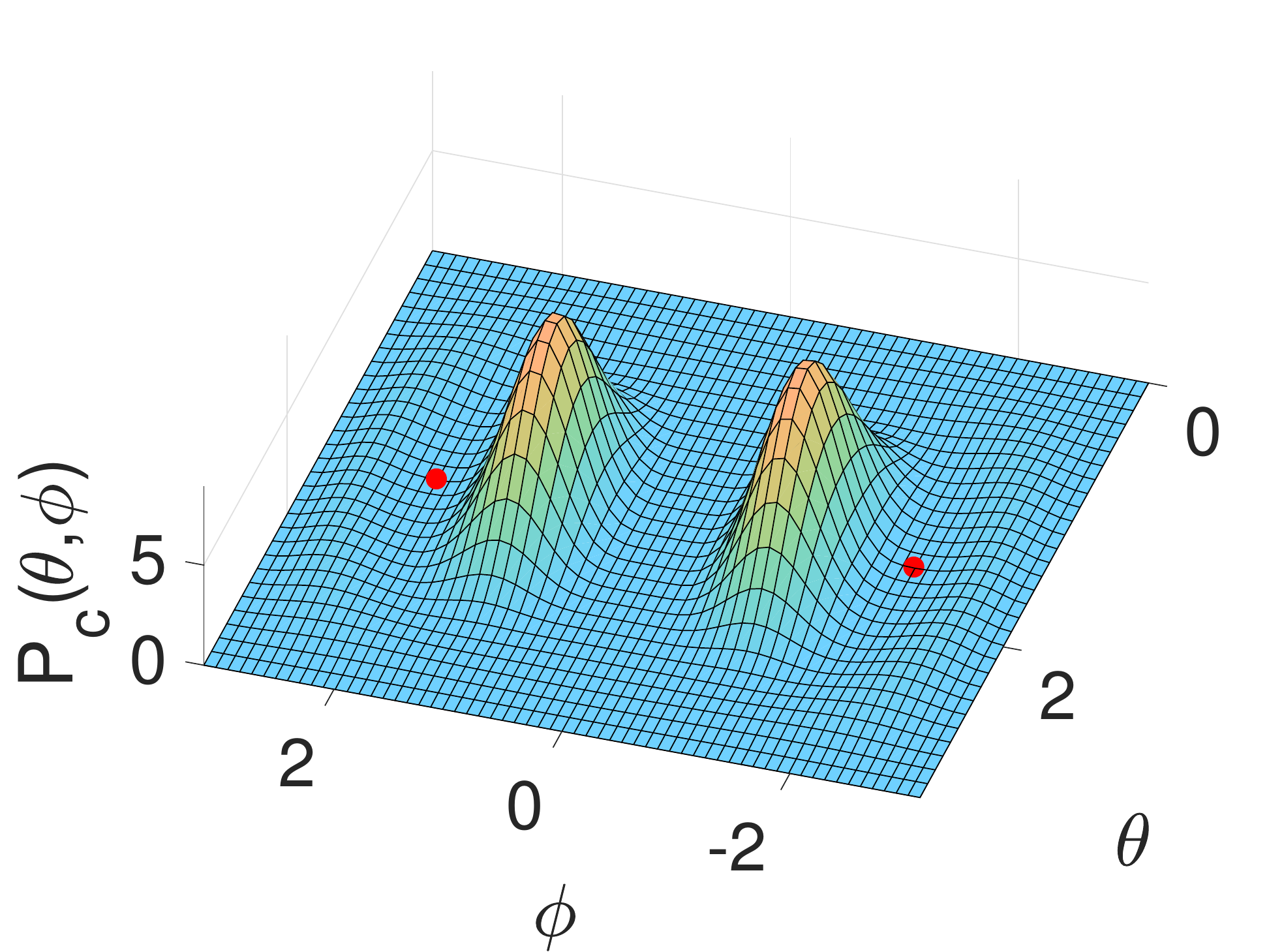}}\hspace{0.1 in}
\subfigure[\ $\Psi^{A_e}$, $R=100$]{ \includegraphics[width=0.22\textwidth]{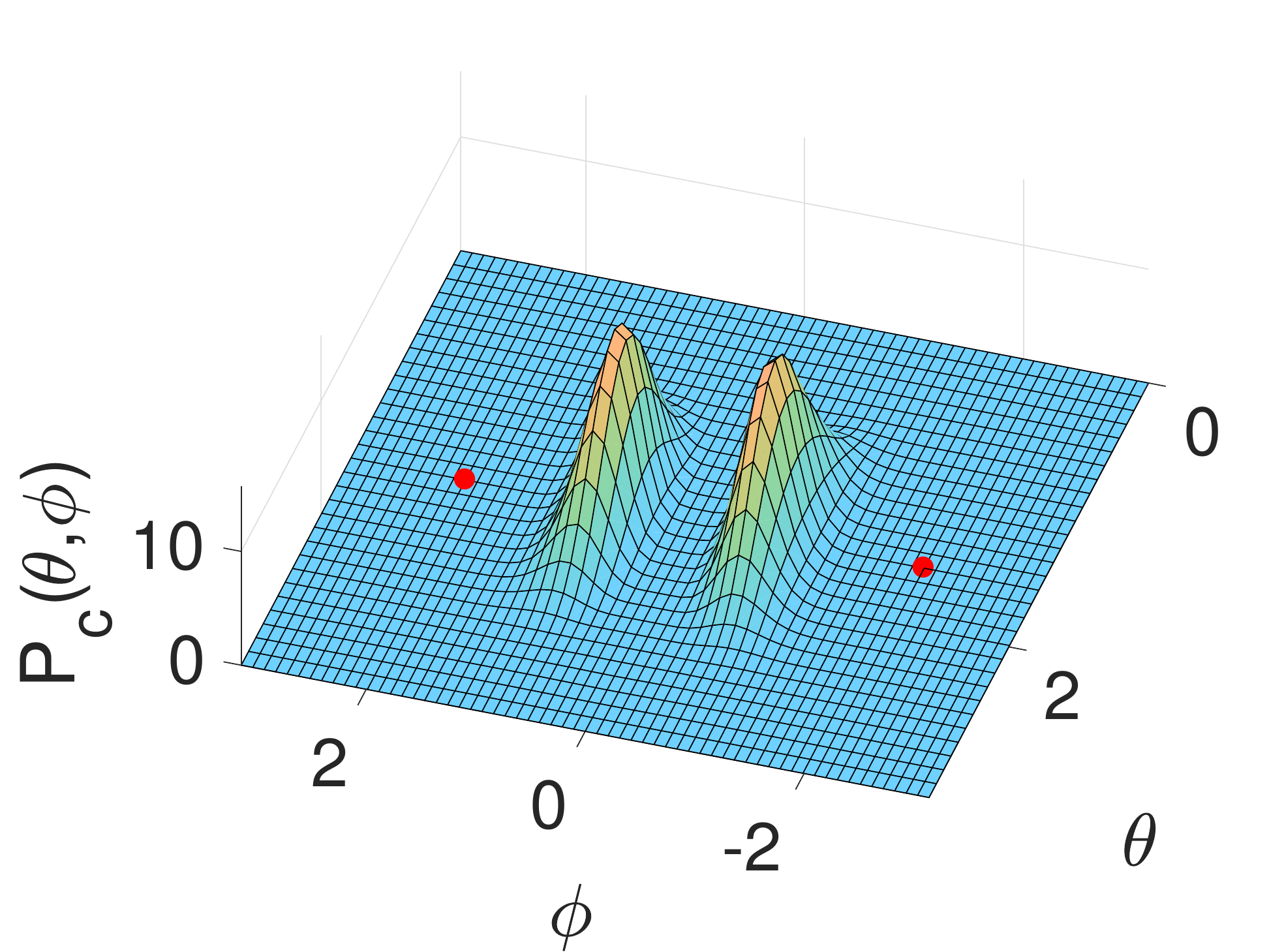}} \hspace{0.1 in}
\subfigure[\ $\Psi^{A_e}$, $R=1000$]{ \includegraphics[width=0.22\textwidth]{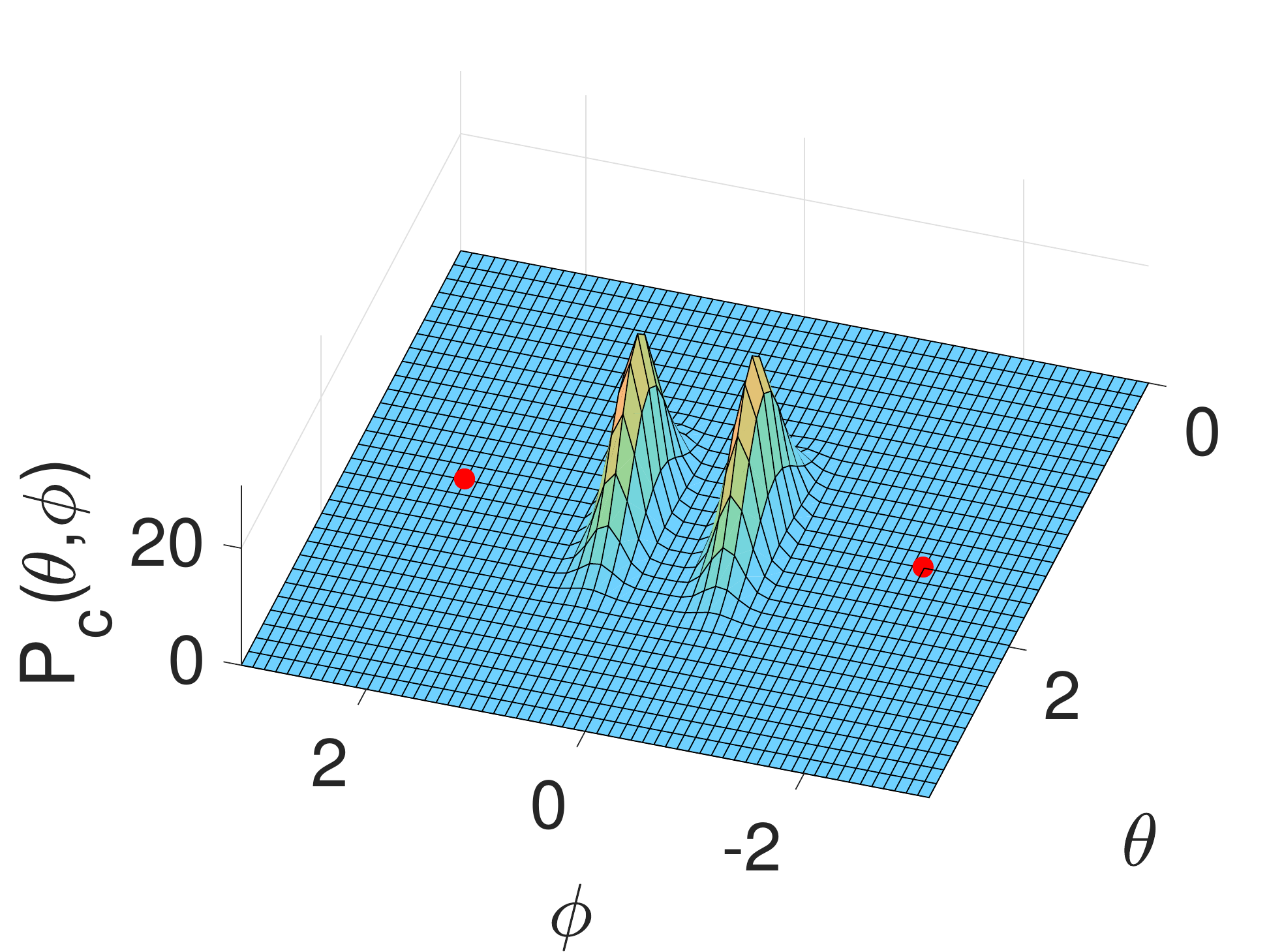}} \hspace{0.1 in}
   \caption{Probability density distribution $P_c(\theta,\phi)$ of any electron when the
other two electrons are fixed at two vertices of a regular triangle
circumscribed by the equator of the sphere. The fixed electrons are indicated by the red dots. The in-plane and out-of-plane vibrations of the electrons in the $\Psi^{A_o}$
and $\Psi^{A_e}$ states are characterized by $\{n_1=1,\ n_2=0\}$ and $\{n_1=0,\
n_2=3\}$, respectively. See
text for more information. }
   \label{three_cross}
\end{figure*}

In Fig.~\ref{three_rho}, we show the distribution of the probability density
$\rho_3(\gamma)$ for both $\Psi^{A_o}(\vec{r}_1,\vec{r}_2,\vec{r}_3)$ and
$\Psi^{A_e}(\vec{r}_1,\vec{r}_2,\vec{r}_3)$. $\rho_3(\gamma)$ is the probability of finding any two of the three
electrons with angular separation $\gamma$. $\rho_3(\gamma) = 8\pi^2R^4 P(\vec{r}_1,\vec{r}_2)$, where
$P(\vec{r}_1,\vec{r}_2)=\int P(\vec{r}_1,\vec{r}_2,\vec{r}_3)dS_3$,
and $P(\vec{r}_1,\vec{r}_2,\vec{r}_3)=|\Psi^{A_p}(\vec{r}_1,\vec{r}_2,\vec{r}_3)|^2$. 
The factor $8\pi^2$ arises from the normalization of $\rho_3(\gamma)$. From
Fig.~\ref{three_rho}(a) for the case of $\Psi^{A_o}$, with the increase of $R$, we see the movement of the
peak towards $\gamma=2.1 \approx 2\pi/3$ and, simultaneously, the shrinking
width of the peak. The value of $2\pi/3$ for $\gamma$ is recognized as the angular distance between any neighboring vertices
in a triangular configuration of electrons on the equator. It signifies the enhanced correlation and localization of the
electrons with $R$. In contrast, the $\Psi^{A_e}$ state exhibits distinct behaviors. From
Fig.~\ref{three_rho}(b), we see three peaks
on the curve of $R=1000$. These peaks correspond to
distinct vibration modes, which will be discussed in the next section.

We also define the 
mean inverse separation $\tilde{d}_{ee}^{-1}$ of any two electrons to
characterize their correlation. $\tilde{d}_{ee}^{-1}
=\int\int\frac{R}{|\vec{r}_i-\vec{r}_j|}P(\vec{r}_i,\vec{r}_j)dS_idS_j$.
We recognize that $\tilde{d}_{ee}^{-1}=V_0R/3$. It is numerically shown that
$\tilde{d}_{ee}^{-1}$ decreases monotonously with $R$ and approaches $1/\sqrt{3} \approx 0.577$ in the
large-$R$ limit. Specifically, for $R$ increasing from 5000 to
10 000, $\tilde{d}_{ee}^{-1}$ decreases slightly from $0.582$ to $0.581$ for $\Psi^{A_o}$ and from $0.586$
to $0.584$ for $\Psi^{A_e}$.

{\bf{Asymptotic behaviors in the small- and large-$R$ regimes}} In this section, we present asymptotic
analysis of the ground states of the three-electron system in the small- and large-$R$
regimes. The relevant theoretical results are consistent with the energy curves in
Fig.~\ref{three_energy}.

For small $R$, we perform perturbation analysis around the zeroth order wave function
$\Psi^{A_p}_{\mathbf{L}_0}$, where the subscript refers to the state of
$\mathbf{L}_0=\{l_1=1,l_2=1,l_3=1\}$ for $\Psi^{A_o}$ and
$\mathbf{L}_0=\{l_1=1,l_2=2,l_3=3\}$ for $\Psi^{A_e}$. For both cases,
$l=0$, and $m=0$. The zeroth- and first-order corrections
to the ground-state energy $E_0$ are: $E^{(0)}_0=K_0 =
\left(l_1(l_1+1)+l_2(l_2+1)+l_3(l_3+1)\right)\hbar^2/R^2$, and $E^{(1)}_0=V^{A_p}_{\mathbf{L}_0
\mathbf{L}_0}$. Specifically, $V^{A_o}_{\mathbf{L}_0 \mathbf{L}_0} \approx 2.4/R$,
$V^{A_e}_{\mathbf{L}_0 \mathbf{L}_0} \approx 2.6/R$. Therefore, up to the first-order
correction, we have $V_0/K_0 = E^{(1)}_0/E^{(0)}_0 \propto R$. This linear dependence of
$V_0/K_0$ on
$R$ agrees well with the numerical result presented in the inset of
Fig. \ref{three_energy}(b). For $R \leq 1$, the maximum deviation of $E_0$ from the
CI method and the perturbation analysis up to the first-order correction is less than
$1\%$ (see SI).

We proceed to analyze the small vibration of the three strongly correlated electrons in the
large-$R$ regime. The equilibrium positions of the electrons are at the vertices of a regular triangle circumscribed by the equator of the
sphere: 
$(\bar{\theta}_1 =\pi/2,\bar{\phi}_1=0)$,
$(\bar{\theta}_2=\pi/2,\bar{\phi}_2=2\pi/3)$, and
$(\bar{\theta}_3=\pi/2,\bar{\phi}_3 = 4\pi/3)$. By introducing a set of collective coordinates $\eta_r$
($r=1,2...6$) like in the treatment of the two-electron system, the classical Hamiltonian of the three-electron system is 
\begin{equation}\label{Hos}
H=\frac{1}{2}R^2\sum\limits_{r=1}^6\dot{\eta}_r^2+\frac{1}{2}\sum\limits_{r=1}^3\omega_r^2R^2\eta_r^2+\frac{\sqrt{3}}{R}, 
\end{equation}
where $\omega_1=3^{-\frac{1}{4}}R^{-\frac{3}{2}}$,
$\omega_2=\omega_3=(\sqrt{5}/2)\omega_1$ (see SI). These six collective coordinates describe three types of vibrations: the relative in-plane vibration between any two electrons (by $\eta_1$ and $\eta_2$), the out-of-plane
vibration (by $\eta_3$), and the rotation of the whole system along three mutually
perpendicular axes (by $\eta_4$, $\eta_5$, and $\eta_6$). 
In the large-$R$ regime, the
vibrational energy is proportional to $R^{-3/2}$, and the rotational energy
scales with $R$ in the form of $R^{-2}$ (see SI). By ignoring the rotational motion,
quantization of the Hamiltonian in Eq.~(\ref{Hos}) leads to the following asymptotic
expression for the energy levels in the three-electron system in the
large-$R$ regime:
\begin{equation}
E_{\{n_1,n_{2}\}}=(n_1+\frac{1}{2})\hbar\omega_1+(n_{2}+1)\hbar\omega_2+\frac{\sqrt{3}}{R}.
\label{E3_n1n2}
\end{equation}
Applying the virial theorem to Eq. (\ref{E3_n1n2}), we obtain the asymptotic
expression for $V_0/K_0$:  $\lim_{R\rightarrow \infty}V_0/K_0 \rightarrow
R^{\frac{1}{2}}$. The numerically solved $V_0/K_0$-$R$ curves as shown in
Fig. \ref{three_energy}(b) are in good agreement with this power law.

To characterize the correlation of the three electrons on the sphere, we calculate the
probability density distribution $P_c(\theta,\phi)$ of any electron when the
other two electrons are fixed at
$\vec{r}^\ast_2=(\bar{\theta}_2=\frac{\pi}{2},\bar{\phi}_2=\frac{2\pi}{3})$ and
$\vec{r}^\ast_3=(\bar{\theta}_3=\frac{\pi}{2},\bar{\phi}_3=-\frac{2\pi}{3})$.
$P_c(\theta,\phi)=P(\vec{r},\vec{r}^\ast_2,\vec{r}^\ast_3)/P(\vec{r}^\ast_2,\vec{r}^\ast_3)$,
where $P(\vec{r}^\ast_2,\vec{r}^\ast_3)=\int
P(\vec{r}_1,\vec{r}^\ast_2,\vec{r}^\ast_3)|dS_1$ and
$P(\vec{r},\vec{r}^\ast_2,\vec{r}^\ast_3)=|\Psi^{A_p}(\vec{r},\vec{r}^\ast_2,\vec{r}^\ast_3)|^2$.
A
striking feature in the profiles of $P_c(\theta,\phi)$, as shown in
Fig.~\ref{three_cross}, is the appearance of the
double peaks. For the odd-parity case in Figs.~\ref{three_cross}(a)-\ref{three_cross}(d), 
the peaks near $\theta=0$ are out of the plane of
the equator. In contrast, for the even-parity case shown in
Figs.~\ref{three_cross}(e)-\ref{three_cross}(h), the peaks are in the plane of the equator.
These two classes of ground-state vibration modes are completely determined by the parity
of the wave function.

To determine the values for $n_1$ and
$n_2$ in Eq.(\ref{E3_n1n2}) in the ground states, we compare a series of $V_0/K_0$-$R$ curves using trial values for
$n_1$ and $n_2$ with that from the CI method. It turns out that $n_1=1$, $n_2=0$ for
$\Psi^{A_o}$, and $n_1=0$, $n_2=3$ for $\Psi^{A_e}$. The difference in the vibration modes
is related to the distinct nodal structures caused by the opposite parities
of $\Psi^{A_o}$ and
$\Psi^{A_e}$~\cite{nodal}.  Here, it is of interest to note the
appearance of the peaks in the $P_c(\theta,\phi)$ profiles even at
relatively small $R$, as shown in Fig.~\ref{three_cross}(b)
and~\ref{three_cross}(e). This observation suggests that the vibration modes are determined by the
symmetry of the wave function instead of the size of the system. Increasing $R$ enhances
these pre-existenting vibration modes.

\section{Conclusion}

In summary, we generalized the classical Thomson problem to the quantum regime to explore the
underlying physics in electron correlations. We constructed symmetrized ground-state wave
functions based on the CI method, systematically investigated the energetics and electron
correlations, and proposed a small-oscillation theory to analyze the
collective vibration modes of the electrons. As a key result of this work, we illustrated
the routine to the strongly correlated, highly localized electron states with the
expansion of the sphere. These results provide insights into the manipulation of
electron states by exploiting confinement geometry. 
Finally, it is of interest to speculate on the connection of the $N$-electron system to
the classical Thomson problem~\cite{Bowick2002, Bowick2006}.  Despite the challenge in theory to construct the ground-state wave function of the $N$-electron system, experimentally, the spontaneous convergence of the
electron state to the highly localized configuration with the expansion of the sphere may
lead to a global solution to the 100-year-old, still unsolved
classical Thomson problem~\cite{Bowick2002, loos2011thinking, agboola2015uniform}.

\
\section*{Acknowledgments}

This work was supported by NSFC Grant No. 16Z103010253, the SJTU startup fund
under Grant No. WF220441904, and an award of the Chinese Thousand Talents
Program for Distinguished Young Scholars under Grants No.16Z127060004 and
No. 17Z127060032.

\end{document}